\documentclass[fleqn,usenatbib]{mnras}

\usepackage{newtxtext,newtxmath}
\usepackage[T1]{fontenc}
\usepackage{ae,aecompl}

\usepackage{graphicx}
\usepackage{float}
\usepackage{times}
\usepackage{amsmath}
\usepackage{amssymb}
\usepackage{bm}
\usepackage{color}
\usepackage{xcolor}
\usepackage{rotating}
\usepackage{pdflscape}
\usepackage{multicol}
\usepackage{ulem}

\def\isoni{$^{56}{\rm Ni}$}
\def\isoco{$^{56}{\rm Co}$}
\newcommand{\cmfgen}{{\sc cmfgen}}
\newcommand{\lb}{$\lambda$}
\newcommand{\gray}{$\gamma$-ray}
\newcommand{\grays}{$\gamma$-rays}
\newcommand{\kms}{\hbox{km$\,$s$^{-1}$}}
\newcommand{\Msun}{\hbox{M$_{\odot}$}}
\newcommand{\Mch}{\hbox{M$_{\rm Ch}$}}

\newcommand{\caii}{\mbox{Ca~{\sc ii}}}
\newcommand{\ariii}{\mbox{Ar~{\sc iii}}}

\newcommand{\oi}{\mbox{O~{\sc i}}}

\newcommand{\tkii}{\mbox{Ti~{\sc ii}}}
\newcommand{\fei}{\mbox{Fe~{\sc i}}}
\newcommand{\feii}{\mbox{Fe~{\sc ii}}}
\newcommand{\feiii}{\mbox{Fe~{\sc iii}}}

\newcommand{\coii}{\mbox{Co~{\sc ii}}}
\newcommand{\coiii}{\mbox{Co~{\sc iii}}}

\newcommand{\nkii}{\mbox{Ni~{\sc ii}}}
\newcommand{\nkiii}{\mbox{Ni~{\sc iii}}}
\newcommand{\siii}{\mbox{S~{\sc iii}}}

\newcommand{\iso}[2]{$^{#2}{\rm #1}$}
\newcommand{\ions}[2]{{#1}$^{#2}$}
\newcommand{\mum}{\hbox{$\mu$m}}
\def\myunit#1{\unskip\,\textrm{#1}}
\def\fig{Fig.}
\def\figs{Figs.}

\def\rrte{relativistic radiative transfer equation}

\title[Nebular SNe Ia Spectra]{Understanding nebular spectra of Type Ia supernovae}

\author[Kevin D. Wilk, D. John Hillier, Luc Dessart]
{Kevin D. Wilk,$^{1}$\thanks{E-mail: kdw25@pitt.edu} D. John Hillier,$^1$\thanks{E-mail: hillier@pitt.edu}  Luc Dessart$^2$ \\
$^1$ Department of Physics and Astronomy \& Pittsburgh Particle physics, Astrophysics, and Cosmology Center (PITT PACC), University of Pittsburgh,  \\ Pittsburgh, PA 15260, USA \\
$^2$ Unidad Mixta Internacional Franco-Chilena de Astronom\'ia (CNRS UMI 3386), 
Departamento de Astronom\'ia, Universidad de Chile, \\
Camino El Observatorio 1515, Las Condes, Santiago, Chile}

\date{Accepted XXX. Received YYY; in original form ZZZ}

\pubyear{2019}

\begin{document}
\label{firstpage}
\pagerange{\pageref{firstpage}--\pageref{lastpage}}
\maketitle

% Abstract of the paper
\begin{abstract}
In this study, we present one-dimensional, non-local-thermodynamic-equilibrium, radiative transfer simulations (using {\cmfgen}) in which we introduce micro-clumping at nebular times into two Type Ia supernova ejecta models. We use one sub-Chandrasekhar (sub-\Mch) ejecta with 1.02 \Msun\ and one \Mch\ ejecta model with 1.40 \Msun. We introduce clumping factors $f=0.33,0.25,$ and 0.10 which are constant throughout the ejecta and compared to the unclumped $f=1.0$ case. We find that clumping is a natural mechanism to reduce the ionization of the ejecta, reducing emission from [\feiii], [\ariii], and [\siii] by a factor of a few. For decreasing values of the clumping factor $f$, the [\caii] \lb\lb7291,7324 doublet became a dominant cooling line for our \Mch\ model but still weak in our sub-\Mch\ model. Strong [\caii] \lb\lb7291,7324 indicates non-thermal heating in that region and may constrain explosion modelling. Due to the low abundance of stable nickel, our sub-\Mch\ model never showed the [\nkii] 1.939 micron diagnostic feature for all clumping values.
\end{abstract}

% Select between one and six entries from the list of approved keywords.
% Don't make up new ones.

%%%%%%%%%%%%%%%%%%%%%%%%%%%%%%%%%%%%%%%%%%%%%%%%%%

%%%%%%%%%%%%%%%%% BODY OF PAPER %%%%%%%%%%%%%%%%%%

\section{Introduction}
%Our best understanding of Type Ia supernovae (SNe Ia) is they are thermonuclear explosions of carbon-oxygen (C/O) white dwarfs (WDs) \citep{Hoyle1960}.
The general consensus is that Type Ia supernovae (SNe Ia) are thermonuclear explosions of carbon-oxygen (C/O) white dwarfs (WDs) \citep{Hoyle1960}. Whether this explosion is the result of a system of one WD and a non-degenerate star (known as the single degenerate (SD) channel) or via a system of two WDs (known as the double degenerate (DD) channel) remains uncertain. 

%SNe Ia come from compact WDs and would otherwise cool quickly via adiabatic expansion without additional energy being deposited, so what powers the luminosity of SNe Ia is the decays of radioactive material produced during the explosion.
SNe Ia come from compact WDs and cool quickly via adiabatic expansion, and without an additional energy supply, they would be extremely difficult to detect. What powers the observed luminosity of SNe Ia is the decay of radioactive material produced during the explosion. The main radioactive isotope produced is \isoni, whose decay chain is \isoni$\rightarrow$\isoco$\rightarrow$\iso{Fe}{56}, releasing 1.72 and 3.75 MeV for each part of the decay chain. Therefore, the production of \isoni\ is important in powering the luminosity of SNe Ia. However, the nickel yields (both stable and unstable) in SNe Ia are sensitive to both progenitor mass ($\rho_c$) and explosion scenario.

%In 1D explosion modelling, neutrons are more easily captured during nuclear burning in higher central densities. 
In 1D explosion modelling, higher central densities lead to enhanced electron capture and thus a larger neutron excess during the explosion.
As a consequence, more stable nickel (\iso{Ni}{58}, \iso{Ni}{60}, and \iso{Ni}{62}) is produced \citep{Nomoto1984,Khokhlov1991a,Khokhlov1991b}. Sub-\Mch\ WDs have lower central densities, and 1D modelling of SNe Ia from sub-\Mch\ progenitors shows a lower abundance of \iso{Ni}{58} and \iso{Ni}{60} compared to \Mch\ SNe Ia. 
%{\red However, 3D DDT modelling suggests that the \iso{Ni}{56} hole predicted in 1D \Mch\ WD DDT models may be absent, and both \iso{Ni}{56} and \iso{Ni}{58} extend from the lowest velocities to about 10\,000 \kms \citep{Kasen2009,Seitenzahl2013}.}
However, 3D DDT modelling does not produce a \iso{Ni}{56} hole. Instead, the abundance of both \iso{Ni}{56} and \iso{Ni}{58} extend from the lowest velocities to about 10\,000 \kms\ \citep{Kasen2009,Seitenzahl2013}. 
This result arises because ignition occurs in the centre of the WD. If ignition occurs on the surface, as in a double detonation, the burning front moves in and there is no mixing of stable Ni outwards \citep{Woosley1994,Livne1995, Fink2007,Fink2010}. Overall, the 3D simulations of WD explosions remain somewhat artificial, and the outcome depends strongly on number of ignition points and their distribution.
%Despite \iso{Ne}{22} settling in sub-\Mch\ being proposed as a way to enhance the neutronization, the time-scale for gravitational settling can be $\sim10^9-10^{10}$ yrs \citep{Bildsten2001}. 
Despite the time-scale for gravitational settling being $\sim10^9-10^{10}$ yrs \citep{Bildsten2001}, \iso{Ne}{22} settling in sub-\Mch\ is proposed as a way to enhance the neutronization. 
Therefore, nebular nickel and IGE spectral features may constrain the physics of SNe Ia \citep{Woosley1997,Iwamoto1999,Stehle2005,Mazzali2006,Gerardy2007,Maeda2010,Mazzali2011,Mazzali2012,Mazzali2015}.

%We fundamentally do not understand the amount of stable nickel produced, the overall abundances of IMEs, nor where in the ejecta these IMEs are produced relative to the original \isoni. However, studying nebular spectra will allow us to estimate these uncertainties. At nebular times any nickel emission features are due to the remaining stable nickel, particularly from \iso{Ni}{58} and \iso{Ni}{60}. Since the width of any observed nebular feature is influenced by emission over the velocities at which the species exist, nickel features will test model predictions of a \iso{Ni}{56} hole. 
Without knowing the progenitor system and explosion scenario, we fundamentally cannot accurately predict (despite understanding flame physics) the amount of stable nickel produced, the overall abundances of IMEs, nor where in the ejecta these IMEs are produced relative to the \isoni. However, studying nebular spectra will allow us to estimate these properties. At nebular times any nickel emission features are due to the remaining stable nickel, particularly from \iso{Ni}{58} and \iso{Ni}{60}. Since the width of any observed nebular feature is influenced by emission over the velocities at which the species exist, stable nickel features may help constrain the presence of the \isoni\ hole (irrespective of the model).

Nebular spectra are great tools to understand progenitors of SNe Ia. At this time, the ejecta is optically thin to continuum and most lines (with exceptions such as UV transitions), and much of the spectra comes from the inner part of the ejecta ($\lesssim{8000}$ \kms), where the densities are highest and iron is the most abundant species. Because iron is most abundant in this region at nebular times, optical spectra are dominated by \feii\ and \feiii\ lines and exhibit little to no flux from IME species such as \siii\ and \ariii. The [\caii] \lb\lb7291, 7324 may be blended with the [\feii] \lb\lb7155, 7172, 7388 triplet, so its presence is difficult to determine. 

Numerous previous studies have investigated SN Ia nebular spectra \citep{Houck1992,Ruiz-Lapuente1992,Ruiz-Lapuente1995,Smareglia1997,Mazzali1998,Gerardy2002PhD,Kozma2005,Maeda2010a,Blondin2012,Taubenberger2013b,Mazzali2015,Black2016,Botyanszki2017,Graham2017,Maguire2018,Black2018PhD,Diamond2018,Black2018}. Authors often study emission lines by fitting Gaussian profiles to features that may or may not be blended. \cite{Maguire2018}, for instance, fit Gaussian profiles to emission lines, and assumes the levels are in LTE with respect to the ground state. These authors also try and fit the complicated feature around 7300 \AA\ without knowing the possible contribution from [\caii] \lb\lb7291, 7324. Work by \cite{Taubenberger2013b} utilised nebular spectra to understand the emission from [\oi] \lb\lb6300, 6364 in the subluminous SN2010lp (SN1991bg-like). These authors argued for a non-spherical distribution of oxygen located close to the core to produce these features.

\cite{Ruiz-Lapuente1995} struggled to obtain good model fits to nebular spectra despite their models matching the photospheric phase spectra. These authors also note the dominant form of iron is \ions{Fe}{2+}, with a sizeable fraction of \ions{Fe}{+} that need not be coincident with \ions{Fe}{2+}. \cite{Ruiz-Lapuente1992} modelled spectra for distance determination by solving for the ionization structure, assuming collisional excitations dominate, and the energy loss is balanced by the thermalization of \grays\ and positrons from nuclear decays. These authors were able to fit some features (like the [\feiii] 4700 \AA\ blend) to nebular spectra. \cite{Mazzali2015} also obtained good nebular spectral fits to SN2011fe with their `$\rho$-11fe' model by means of abundance tomography. The authors claim that SN2011fe requires the innermost ejecta to be dominated by stable iron, which aids in cooling instead of heating (via radioactive decay) and rules out a low mass ($\sim$1.02 \Msun) WD progenitor. \cite{Mazzali2018} also used abundance tomography to model the fast declining SNe Ia, SN2007bo and SN2011iv. By analyzing emission components of many [\feii] and [\feiii] features, the authors reproduce the spectra by using a two component emission model (one blueshifted and one redshifted), which acts like two distinct nebulae. The authors do not rule out the possibility of an off-centre ignition instead of two colliding WDs. In all of these works, however, the abundances and their distributions are free parameters.

Nebular modelling raises questions about the ionization/abundance structure of ``normal" SN Ia ejecta. Nebular modelling by \cite{Botyanszki2017} and \cite{Wilk2018} predicts strong emission lines of [\siii] \lb\lb9069, 9531 (and [\ariii] \lb71336, 7751 by \cite{Wilk2018}) in their ejecta mass models. Why are these emission lines largely absent or weak in observations? What is the contribution of [\caii] \lb\lb7291, 7324 to the observed 7300 \AA\ feature? What does it imply if IMEs are not seen in nebular spectra? Does this reflect the ionization structure or the abundances and/or chemical stratification? What causes the strength of the [\feiii] 4700 \AA\ feature to be much stronger than other features beyond 5500 \AA\ in models compared to observations \citep{Botyanszki2017,Wilk2018}?

The nebular model spectra of \cite{Wilk2018} indicate a higher ionization than what is generally observed. As clumping enhances the density, increases the recombination rate, and lowers the ionization, we introduce clumping in our nebular modelling of SNe Ia ejecta. Given the high ionization of model SUB1 (a sub-\Mch\ detonation ejecta model from \cite{Wilk2018} -- also see Section~\ref{ch4:section_technique}) at nebular times (because the lack of a ``\iso{Ni}{56} hole'' facilitates more heating of the inner region), clumping is a natural choice given previous evidence of its role in reproducing spectral features \citep{Chugai1992,Bowers1997,Thomas2002,Leonard2005,Leloudas2009,Srivastav2016,Porter2016}. Previous theoretical modelling suggests clumping to be a byproduct of ``nickel bubbles," an expansion of the iron and nickel regions relative to the surrounding material due to radioactive decay energy deposition \citep{Woosley1988,Li1993,Basko1994,Wang2008}, Rayleigh Taylor instabilities during DDT burning \citep{Golombek2005}, or material interaction during detonation \citep{Maier2006}.

%In this paper, we present our work determining the impact of clumping on SNe Ia nebular spectra (+200 d) using model SUB1 and CHAN from \cite{Wilk2018}. 
In this Section, we study the formation of nebular spectra and examine the influence of clumping. We highlight problems with the emission from IMEs and the ionization structure by examining the influence of clumping. In this study, we use models SUB1 and CHAN from \cite{Wilk2018}.
In section~\ref{ch4:section_technique} we discuss our technique for introducing clumping in our models. In section~\ref{ch4:section_results} we present the impact of clumping on nebular SN Ia spectra. We highlight the changes to the ionization structure in section~\ref{ch4:subsection_ionization}. Shifts in ionization are reflected in some species, so in sections~\ref{ch4:subsection_iron},~\ref{ch4:subsection_nickel_line}, and~\ref{ch4:subsection_IMEs} we discuss the effects on iron features, nickel features, and IMEs respectively. Finally, we summarize our work in section~\ref{ch4:section_conclusion}.
%
%88888888888888888888888888888888888888888888888888888888
%
\section{Technique}\label{ch4:section_technique}
%
%88888888888888888888888888888888888888888888888888888888
%
\subsection{Ejecta Models}
This research uses two hydrodynamic models of \cite{Wilk2018}, DDC10 (CHAN) and SCH5p5 (SUB1). DDC10 is a \Mch\ (1.40 \Msun) model from \cite{Blondin2013}. Model SCH5p5 is a sub-\Mch\ (1.04 \Msun) from \cite{Blondin2017}, but we have scaled the density by 0.98 \cite[see][]{Wilk2018} in order to have the same \isoni. Both CHAN and SUB1 have the same \iso{Ni}{56}\ at 0.62 \Msun. We use \cmfgen\ to solve the spherically symmetric, time-dependent, \rrte\ allowing for non-local thermodynamic equilibrium (non-LTE) processes. We take these two models at 216.5 days post-explosion from \cite{Wilk2018}. Table~\ref{ch4:model_info_abund} lists the initial masses for each model as well as the mass abundance of calcium, iron, cobalt, \iso{Ni}{58} plus \iso{Ni}{60}, and \iso{Ni}{56}. We see CHAN has more than twice the amount of stable nickel -- M(\iso{Ni}{58}) + M(\iso{Ni}{60}) -- than SUB1 as well as almost a factor of two more calcium.
%
%88888888888888888888888888888888888888888888888888888888
%
\subsection{Numerical Treatment}\label{ch4:subsection_numerical}
%
%88888888888888888888888888888888888888888888888888888888
%
Our original radiative-transfer calculations for our models did not include the effects of clumping. Therefore, to treat clumping in our models, we make some simple assumptions with \cmfgen. We assume there is no inter-clump media and the clumping is uniform in a homogeneous flow (i.e. $f(V)=f_{_{\rm o}}$ for all species and velocities $V$). Assumptions underlying our clumping approach and discussions of the influence of clumping on core-collapse SNe II-P are provided by \cite{Dessart2018}. Our treatment of clumping in \cmfgen\ differs from that of a concentric shell-type structure, which is susceptible to radiative transfer effects across a shell and requires a large number of depth points to resolve the shells.% This has been implemented for core-collapse SNe II-P \citep{Dessart2018}.

The clumping factor scales many variables, such as densities (scaled by $1/f$) and the emissivities and opacities (calculated using populations derived from clumps then scaled by $f$). This micro-clumping formalism leaves the mass column density unchanged. \cmfgen\ has incorporated this clumping method since \cite{Hillier1999} first applied it to massive stars.

At a time step of 216.5 d post explosion ($\sim$+200 d post maximum) we resolved the \rrte\ for our models SUB1 and CHAN using a clumping factor ($f$) of 0.33, 0.25, and 0.1 (a value motivated by modelling of Wolf-Rayet stars \citep{Hillier1999}). Since the previous time-step was not computed using clumping in the models, we scaled the initial input populations by 1/$f$ between successive ionizations. 
%This simple scaling is adequate since time dependent effects are negligible at this phase.
This simple scaling is adequate since time-dependent effects have a negligible influence on the spectrum.
%
%88888888888888888888888888888888888888888888888888888888
%
\section{Results}\label{ch4:section_results}
%
%88888888888888888888888888888888888888888888888888888888
%
\figs~\ref{ch4:subchan_optical_spec_figure}~and~\ref{ch4:chan_optical_spec_figure} show the synthetic optical and near infrared (NIR) spectra of SUB1 and CHAN for the different values of a clumping factor ($f$ = 1.00, 0.33, 0.25, and 0.10). In order to contrast the little amount of flux in the NIR, we show $\log F_\lambda$ vs. \lb\ for wavelengths 1.0--2.4 $\mu$m. 
%Table~\ref{ch4:table_photometry} contains the calculated photometry for each model.
We also show the component spectrum for $f=0.10$ and $f=0.33$ for models SUB1 and CHAN in \figs~\ref{ch4:sub1_f0p1_spec_components_figure},~\ref{ch4:sub1_f0p33_spec_components_figure},~\ref{ch4:chan_f0p1_spec_components_figure}, and~\ref{ch4:chan_f0p33_spec_components_figure}. 
%
%88888888888888888888888888888888888888888888888888888888
%
\subsection{Unclumped Models}\label{ch4:subsection_unclumped}
%
%88888888888888888888888888888888888888888888888888888888
%
Before we discuss clumping, it is necessary to understand and summarize the results of \cite{Wilk2018} that do not incorporate clumping. This Section focuses on two ejecta models, a direct detonation of a sub-\Mch\ WD and a DDT WD explosion model. Because our models come from two different explosion scenarios, the nucleosynthesis yields and stratification are model dependent. In 1D simulations of DDT explosions in \Mch\ WDs produce what is known as a ``\iso{Ni}{56} hole," which is an underabundance of \iso{Ni}{56} in the innermost region. This arises due to neutron-rich nuclear statistical equilibrium (NSE) burning. Since our SUB1 model has lower central densities during explosion, it does not have a ``\iso{Ni}{56} hole." Therefore, SUB1 has a factor of about 2.26 less stable nickel. Within the ``\iso{Ni}{56} hole'' of CHAN, the temperature and ionization is lower than in the region containing the original \isoni.

%Since stable nickel distribution is centralised to the innermost part of the ejecta, it becomes visible during nebular spectra. Thus, nebular spectra allow us to probe this inner region that was once shielded by the optically thick photosphere.
Since stable nickel is centralized to the innermost part of the ejecta, it only becomes visible as the photosphere recedes inwards and the ejecta becomes optically thin. Hence, the ejecta transitions into the nebular phase. Thus, nebular spectra allow us to probe this once shielded inner region. At 216 days post explosion (roughly +200 days post maximum), we do not see stable nickel in our SUB1 model, unlike in model CHAN. SUB1 has very little stable nickel in the inner region and also has \ions{Ni}{2+} as its dominant ionization (see fig. 7 of \cite{Wilk2018}). At nebular times we expect to see forbidden [\nkii] lines, particularly [\nkii] \lb\lb7378, 7412 and [\nkii] 1.939 \mum, which are present in CHAN. Both models show strong emission from higher ionization states like \ions{Fe}{2+}, \ions{Co}{2+}, \ions{Ar}{2+}, and \ions{S}{2+}.

At nebular times, the energy deposited by radioactive decay is equal to the energy radiated by the gas by numerous cooling lines.
%Our unclumped models tell us non-thermal heating by radioactive decays is heating both the IGEs as well as the IMEs.
In the models, radioactive decay heats both the IGEs as well as the IMEs.
However, it is necessary to determine if these ejecta models will still show spectral signatures of IMEs when clumping is introduced.
%
%88888888888888888888888888888888888888888888888888888888
%
\subsection{Ionization Shifts}\label{ch4:subsection_ionization}
%
%88888888888888888888888888888888888888888888888888888888
%
%As expected, clumping shifts the ionization downward for the IGEs from mostly doubly ionized to singly ionized (i.e. \ions{Fe}{2+}$\rightarrow$\ions{Fe}{+}). The strength of \feiii, \coiii, \nkiii, \siii, and \ariii\ lines decrease considerably with increasing clumping, while \feii\ and \coii\ lines increase. We show the average charge per species for sulfur, argon, calcium, iron, cobalt, and nickel in \figs~\ref{ch4:avg_ion_sub1_figure} and~\ref{ch4:avg_ion_chan_figure}. For both SUB1 and CHAN, the average charge of sulfur, argon, and cobalt differs by almost one electron between $f$ = 1.00 and $f$ = 0.10 in the inner ejecta regions.
We show the average charge per species for sulfur, argon, calcium, iron, cobalt, and nickel for models with different amounts of clumping in \figs~\ref{ch4:avg_ion_sub1_figure} and~\ref{ch4:avg_ion_chan_figure}. As expected, clumping shifts the ionization downward in both models for the IGEs from mostly doubly ionized to singly ionized (i.e. \ions{Fe}{2+}$\rightarrow$\ions{Fe}{+}). The strength of \feiii, \coiii, \nkiii, \siii, and \ariii\ lines decreases considerably with increasing clumping, while \feii\ and \coii\ lines increase. For both SUB1 and CHAN, the average charge of sulfur, argon, and cobalt differs by almost one electron between $f$ = 1.00 and $f$ = 0.10 in the inner ejecta regions.
%
%88888888888888888888888888888888888888888888888888888888
%
\subsection{Impact on Iron Lines}\label{ch4:subsection_iron}
\begin{landscape}
\begin{figure}
\centering
\begin{tabular}{c}
\includegraphics[scale=1, angle=-90]{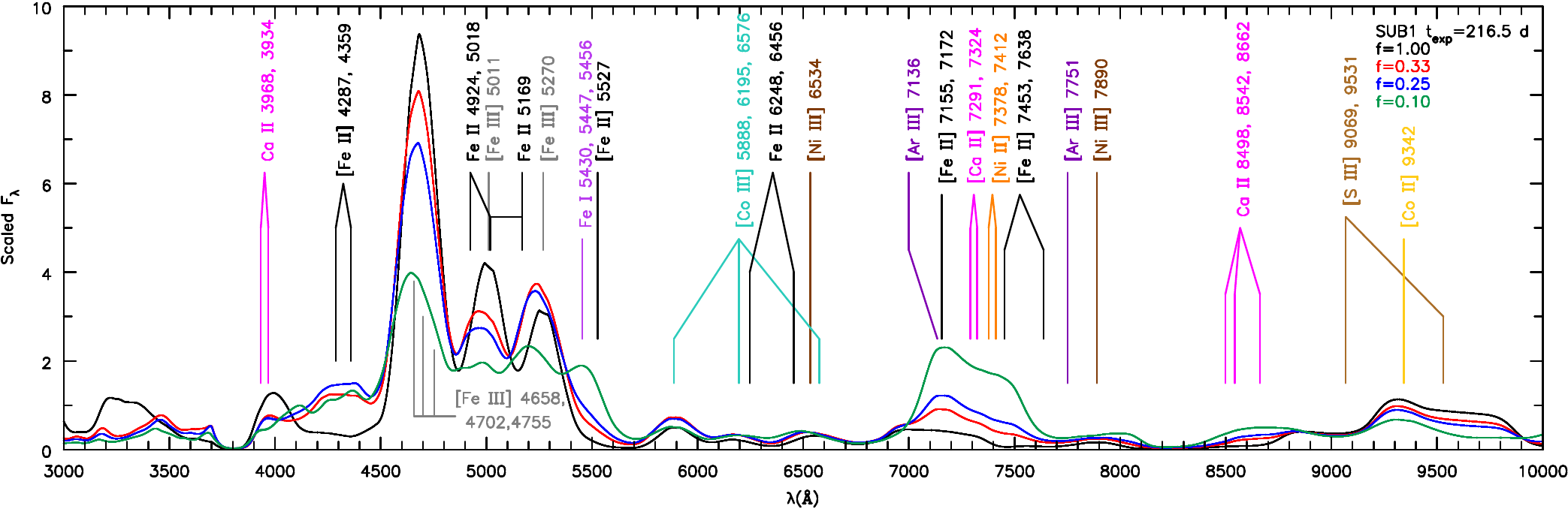} \\
\includegraphics[scale=1, angle=-90]{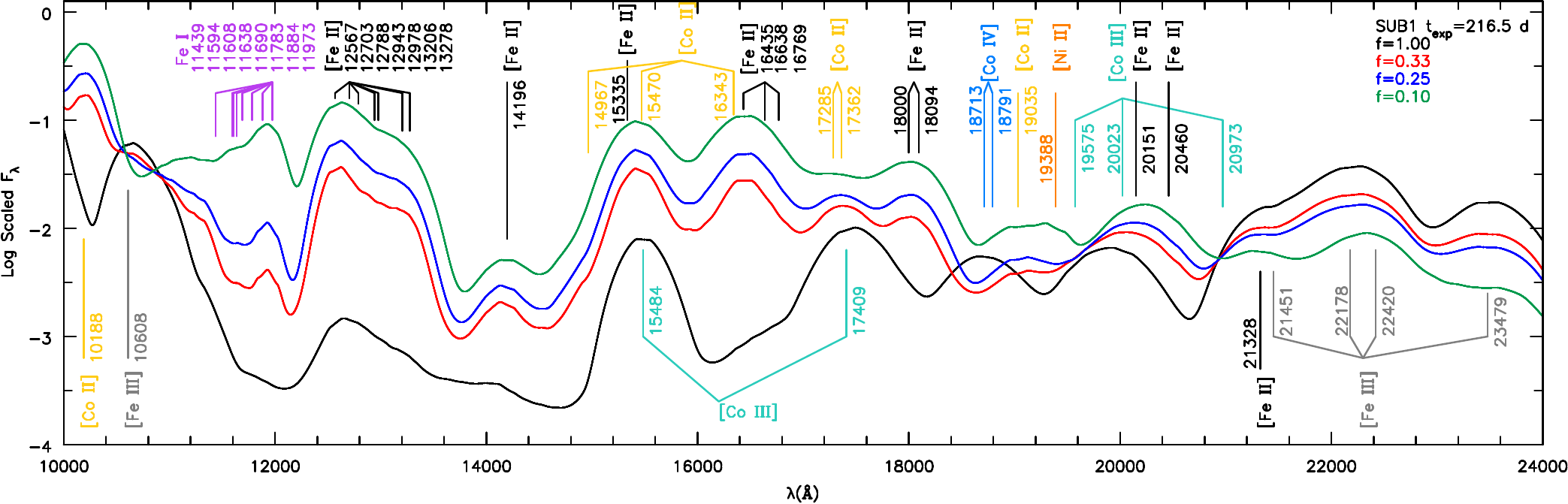}
\end{tabular}
\caption{Spectra of SUB1 at 216.5 days post-explosion compared to the spectra computed using different clumping factor values (0.33, 0.25, and 0.1). The spectra have all been scaled by the same value. To contrast the little flux in the IR, we show the logarithm of the flux. In the IR, particularly from 10\,000-11\,800 \AA, changes are several orders of magnitude for a factor of a few in density.}
\label{ch4:subchan_optical_spec_figure}
\end{figure}
\end{landscape}

\begin{landscape}
\begin{figure}
\centering
\begin{tabular}{c}
\includegraphics[scale=1, angle=-90]{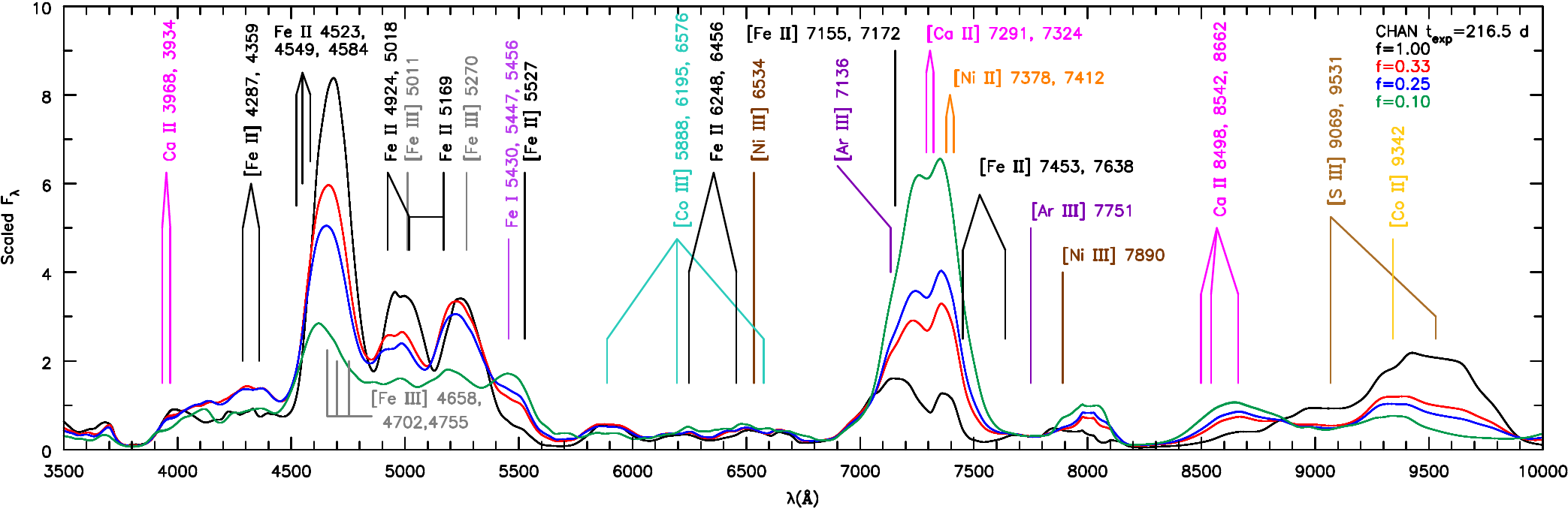} \\
\includegraphics[scale=1, angle=-90]{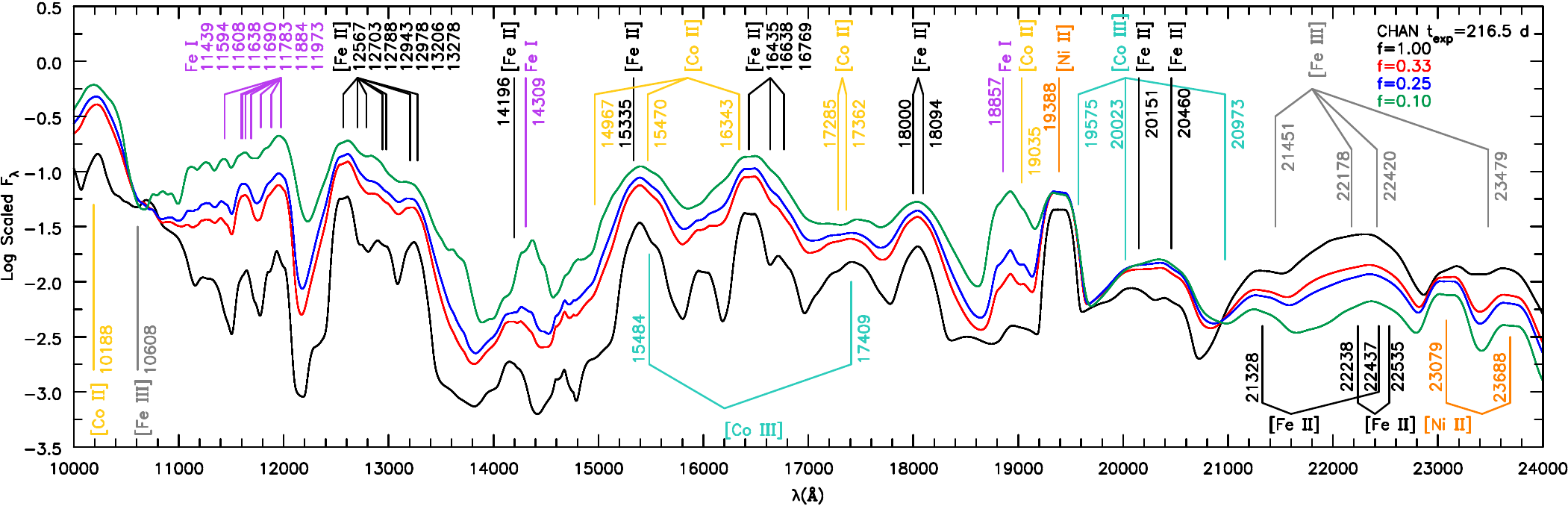}
\end{tabular}
\caption{Spectra of CHAN at 216.5 days post-explosion compared to the spectra computed using different clumping factor values (0.33, 0.25, and 0.1). The spectra have all been scaled by the same value (and also to \fig~\ref{ch4:subchan_optical_spec_figure}). To contrast the little flux in the NIR, we show the logarithm of the flux. Differences between the spectra can be large (greater than a factor of 10) for \fei\ and \feii\ features for just a factor of 3 or more in density. }
\label{ch4:chan_optical_spec_figure}
\end{figure}
\end{landscape}

\begin{figure*}
\centering
\begin{tabular}{c}
\includegraphics[scale=0.75,angle=-90]{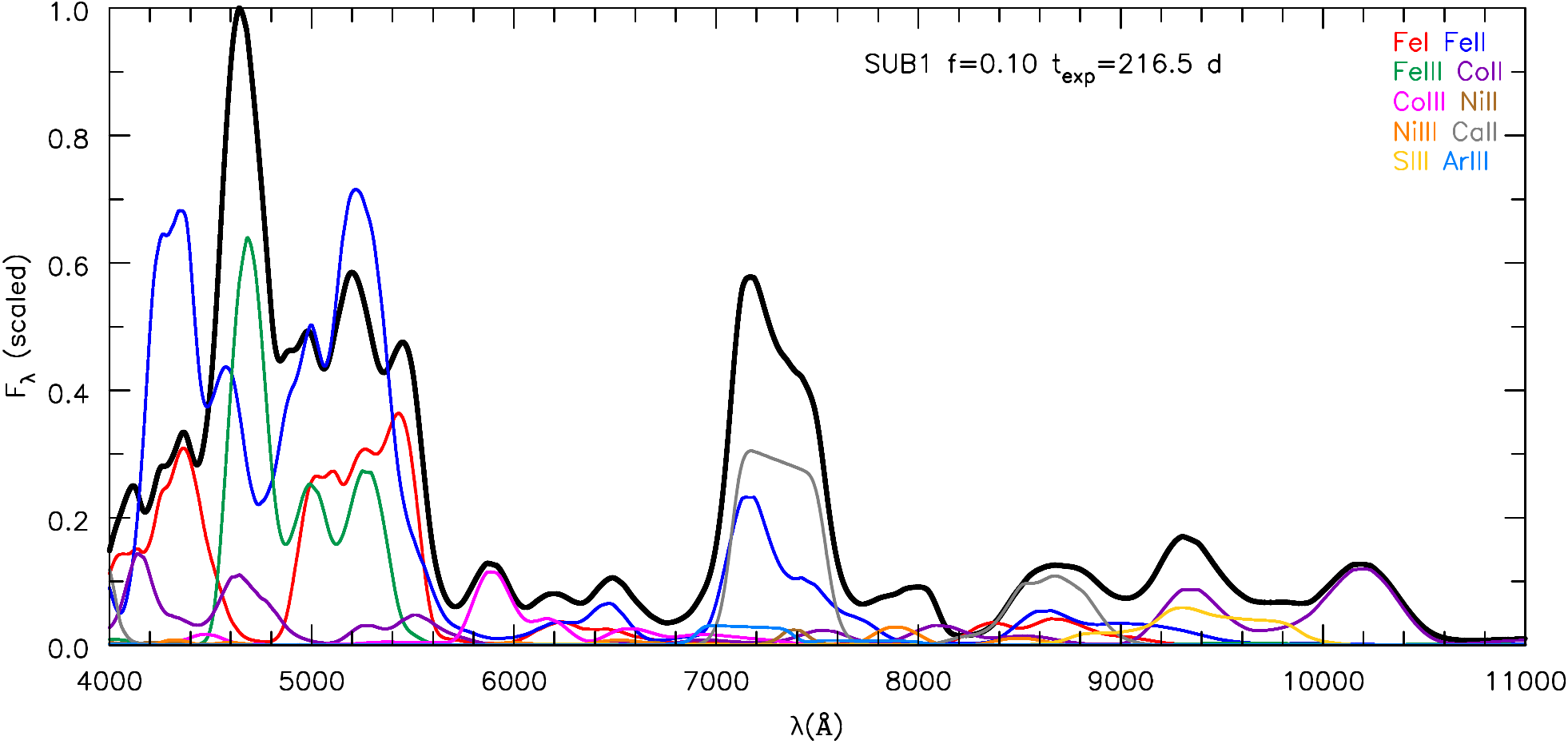} \\
\includegraphics[scale=0.75,angle=-90]{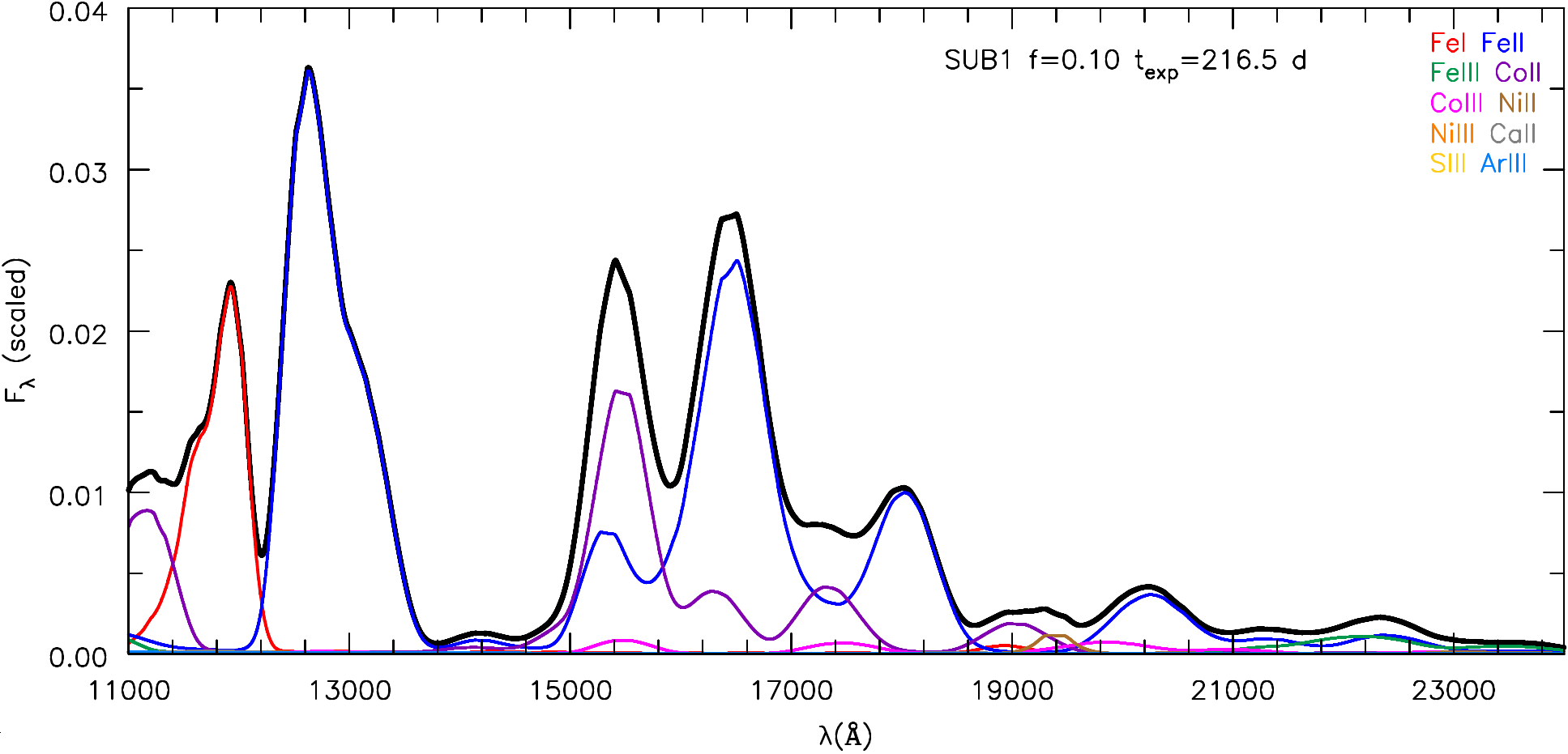}
\end{tabular}
\caption{Observer's frame spectrum (thick black line) of SUB1 along with the component spectra of \fei, \feii, \feiii, \coii, \coiii, \nkii, \nkiii, \caii, \siii, and \ariii\ at 216.5 days post-explosion for a clumping factor of $f=0.10$. 
%We see that the sum of the component spectra produce the total calculated spectrum (thick black line) for wavelengths greater than 6000 \AA since optical depth effects are small or negligible for these wavelengths. 
If the optical depth effects are unimportant, the component spectrum will sum to the total to produce the full spectrum (thick black line). Conversely, we see that the features between 4000-5500 \AA\ cannot be understood without allowing for optical depth effects. 
}
\label{ch4:sub1_f0p1_spec_components_figure}
\end{figure*}

\begin{figure*}
\centering
\begin{tabular}{c}
\includegraphics[scale=0.75,angle=-90]{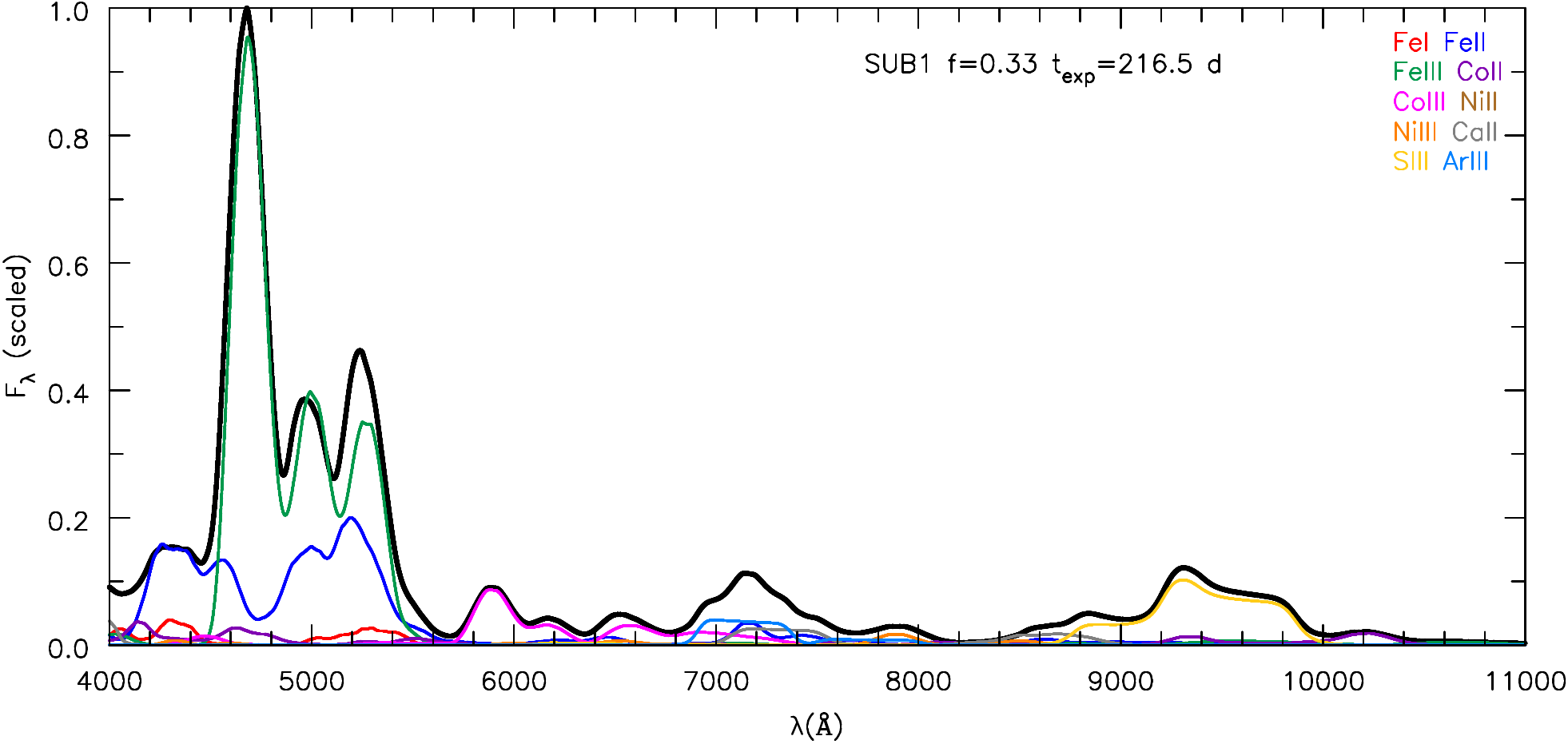} \\
\includegraphics[scale=0.75,angle=-90]{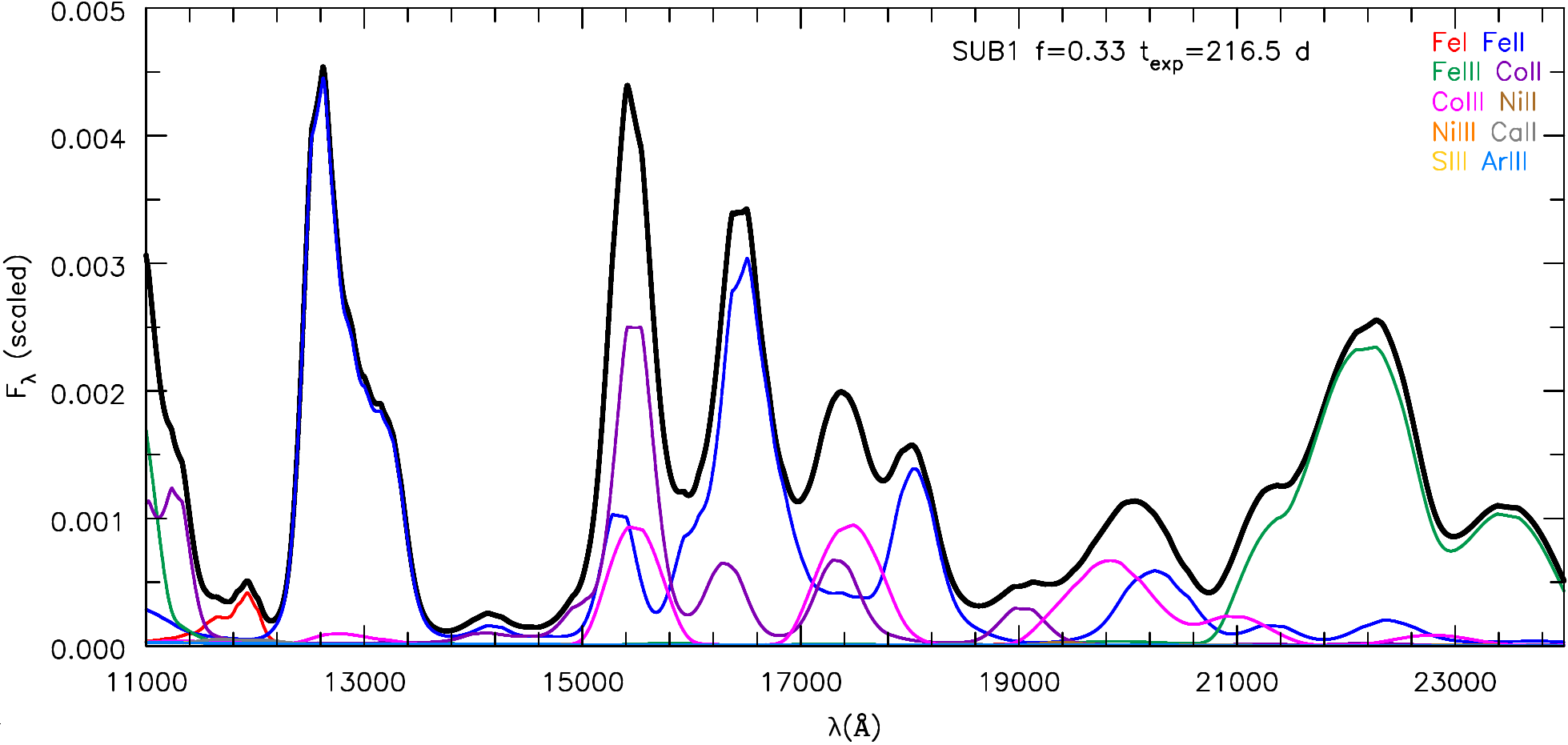}
\end{tabular}
\caption{Observer's frame spectrum (thick black line) of SUB1 along with the component spectra of \fei, \feii, \feiii, \coii, \coiii, \nkii, \nkiii, \caii, \siii, and \ariii\ at 216.5 days post-explosion for a clumping factor of $f=0.33$. 
%We see that the sum of the component spectra produce the total calculated spectrum (thick black line) for wavelengths greater than 6000 \AA since optical depth effects are small or negligible for these wavelengths. 
If the optical depth effects are unimportant, the component spectrum will sum to the total to produce the full spectrum (thick black line). Conversely, we see that the features between 4000-5500 \AA\ cannot be understood without allowing for optical depth effects.}
\label{ch4:sub1_f0p33_spec_components_figure}
\end{figure*}

\begin{figure*}
\centering
\begin{tabular}{c}
\includegraphics[scale=0.75,angle=-90]{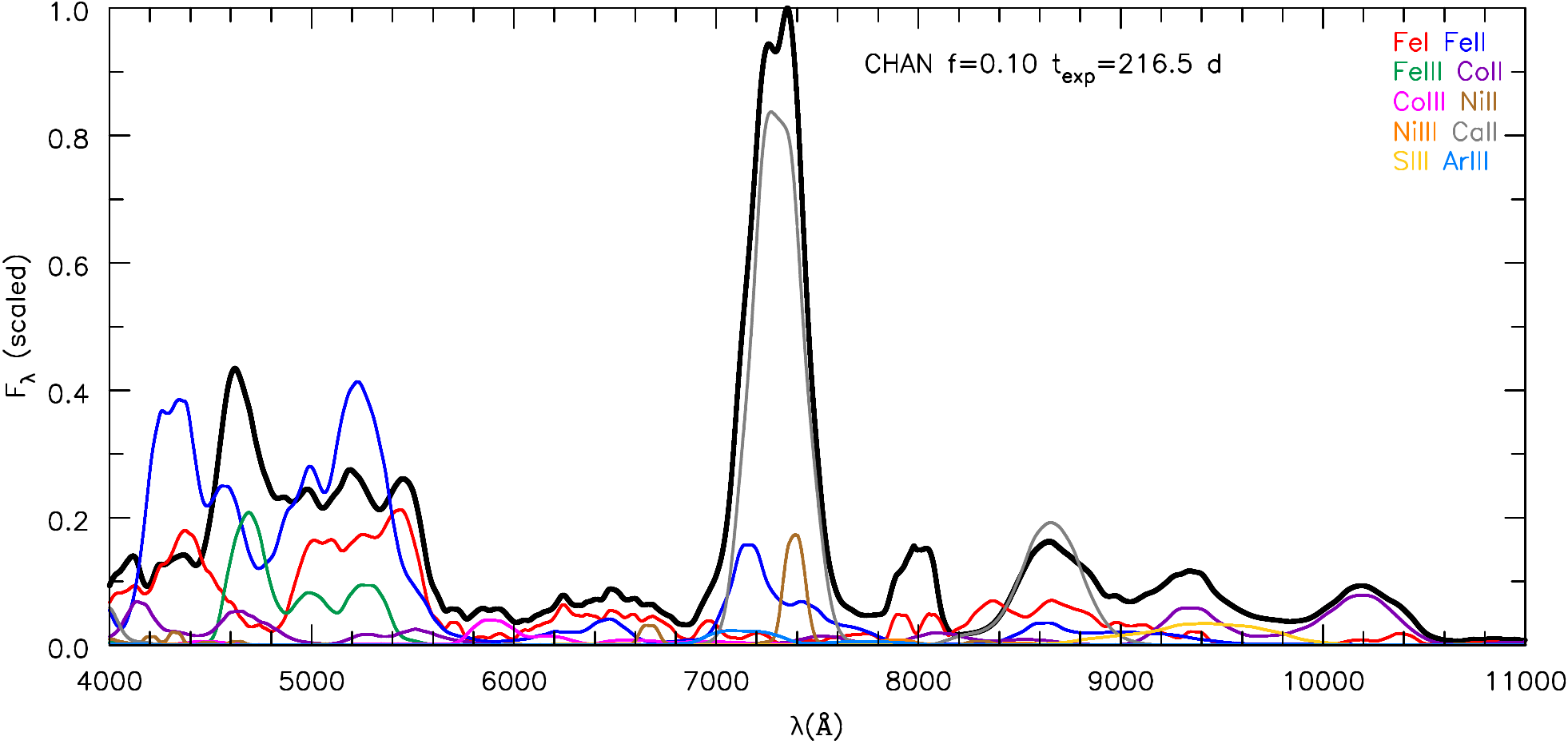} \\
\includegraphics[scale=0.75,angle=-90]{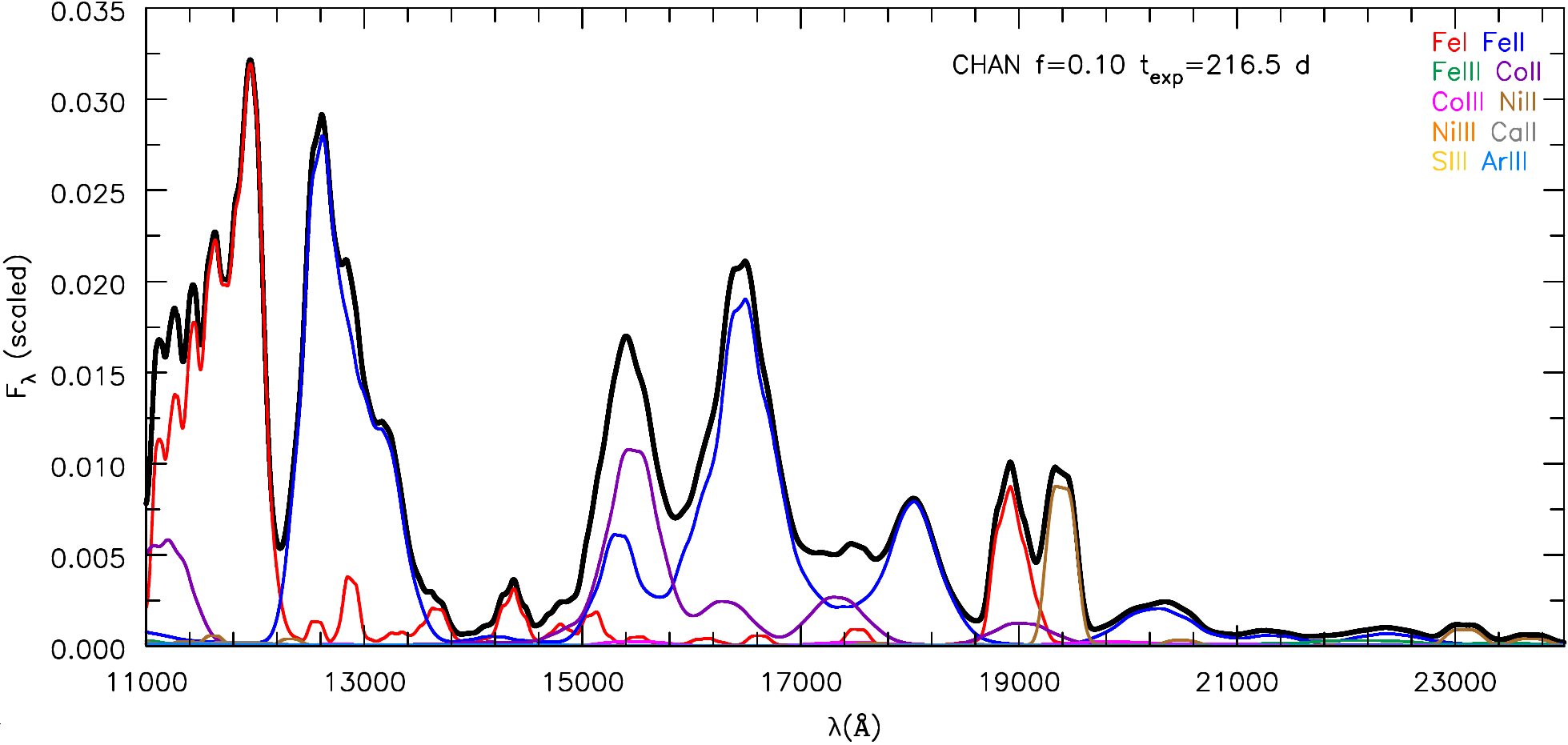}
\end{tabular}
\caption{Observer's frame spectrum (thick black line) of CHAN along with the component spectra of \fei, \feii, \feiii, \coii, \coiii, \nkii, \nkiii, \caii, \siii, and \ariii\ at 216.5 days post-explosion for a clumping factor of $f=0.10$. 
%We see that the sum of the component spectra produce the total calculated spectrum (thick black line) for wavelengths greater than 6000 \AA since optical depth effects are small or negligible for these wavelengths. 
If the optical depth effects are unimportant, the component spectrum will sum to the total to produce the full spectrum (thick black line). Conversely, we see that the features between 4000-5500 \AA\ cannot be understood without allowing for optical depth effects.}
\label{ch4:chan_f0p1_spec_components_figure}
\end{figure*}

\begin{figure*}
\centering
\begin{tabular}{c}
\includegraphics[scale=0.75,angle=-90]{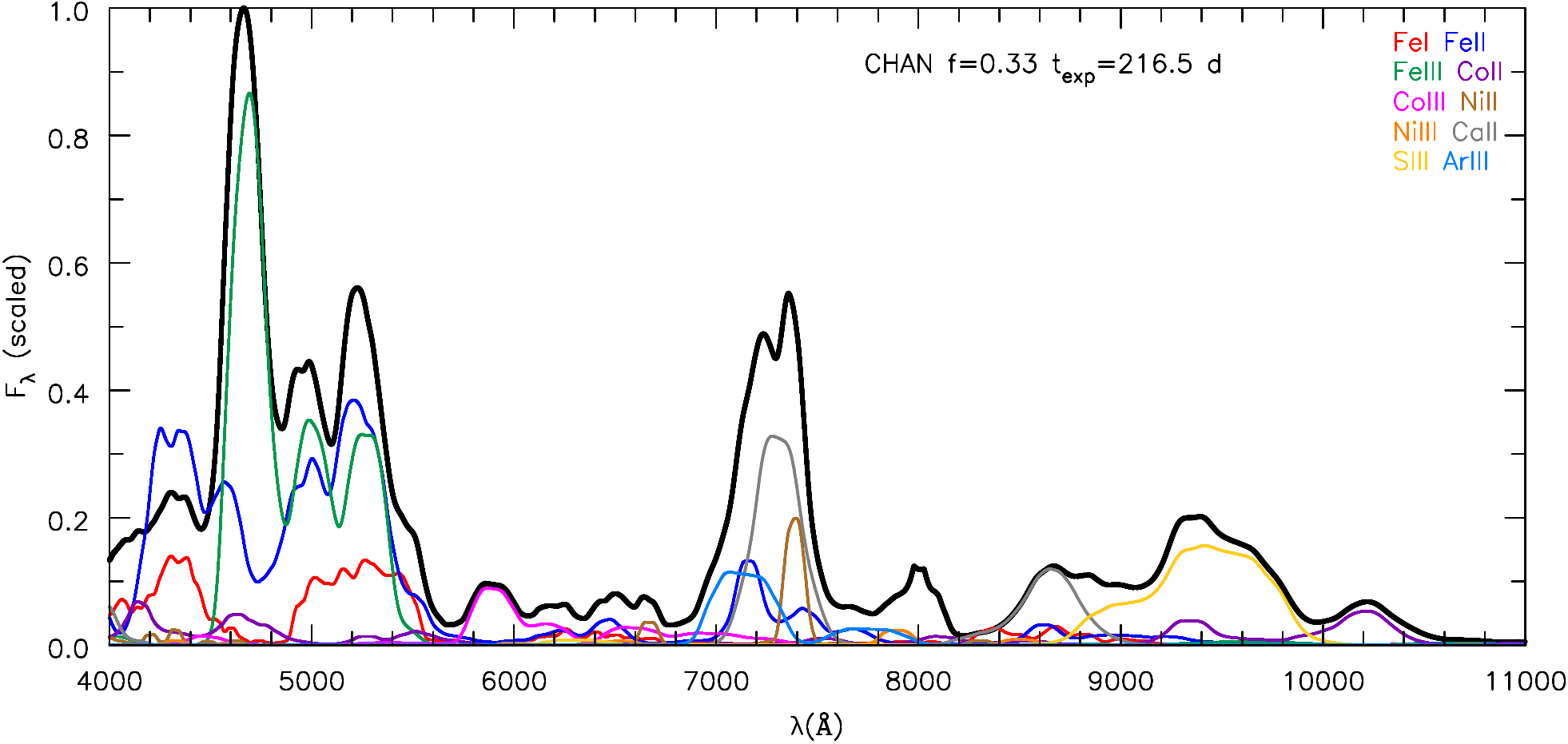} \\
\includegraphics[scale=0.75,angle=-90]{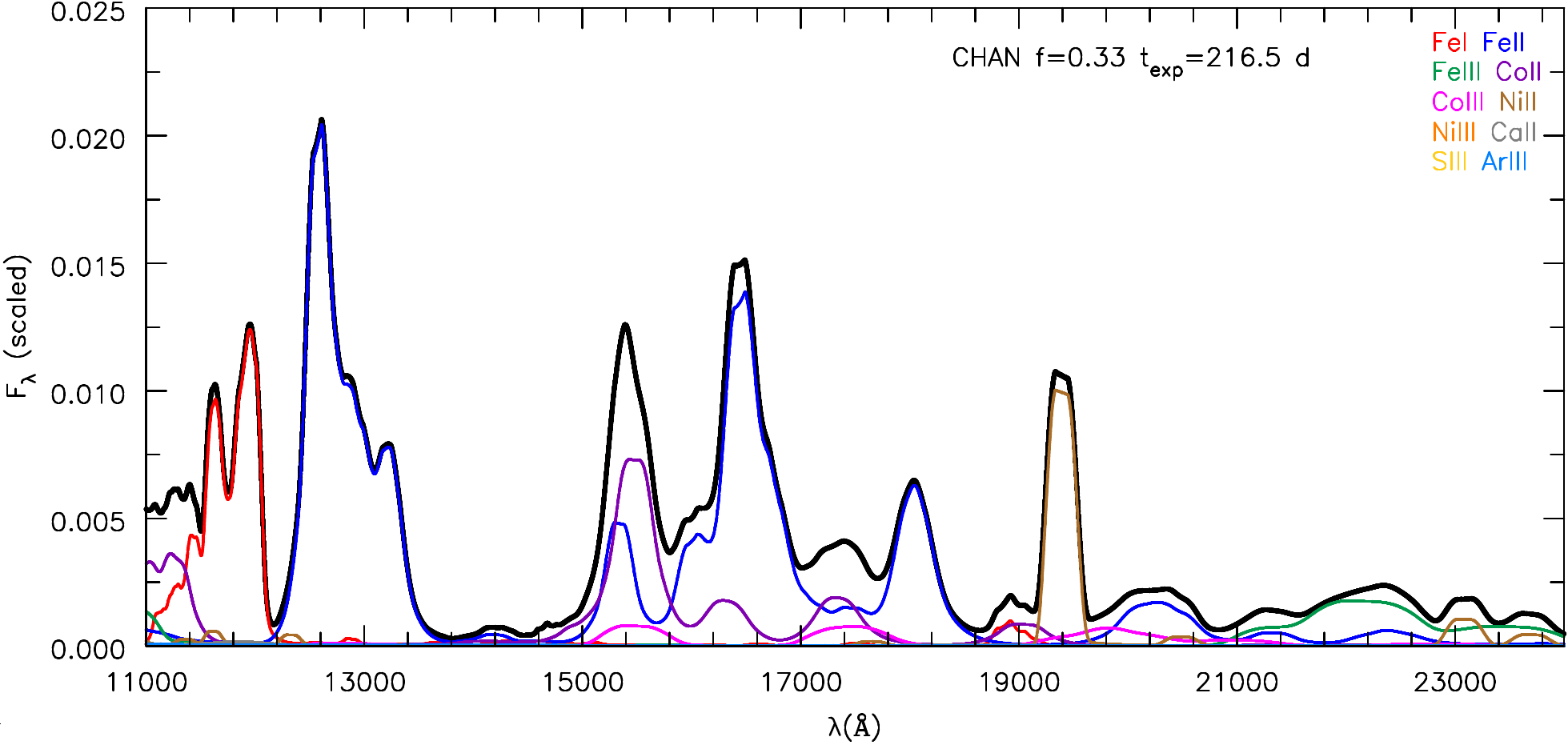}
\end{tabular}
\caption{Observer's frame spectrum (thick black line) of CHAN along with the component spectra of \fei, \feii, \feiii, \coii, \coiii, \nkii, \nkiii, \caii, \siii, and \ariii\ at 216.5 days post-explosion for a clumping factor of $f=0.33$. 
%We see that the sum of the component spectra produce the total calculated spectrum (thick black line) for wavelengths greater than 6000 \AA since optical depth effects are small or negligible for these wavelengths. 
If the optical depth effects are unimportant, the component spectrum will sum to the total to produce the full spectrum (thick black line). Conversely, we see that the features between 4000-5500 \AA\ cannot be understood without allowing for optical depth effects.}
\label{ch4:chan_f0p33_spec_components_figure}
\end{figure*}

\begin{landscape}
\begin{figure}
\begin{minipage}[t]{\linewidth}
\centering
\includegraphics[scale=0.95]{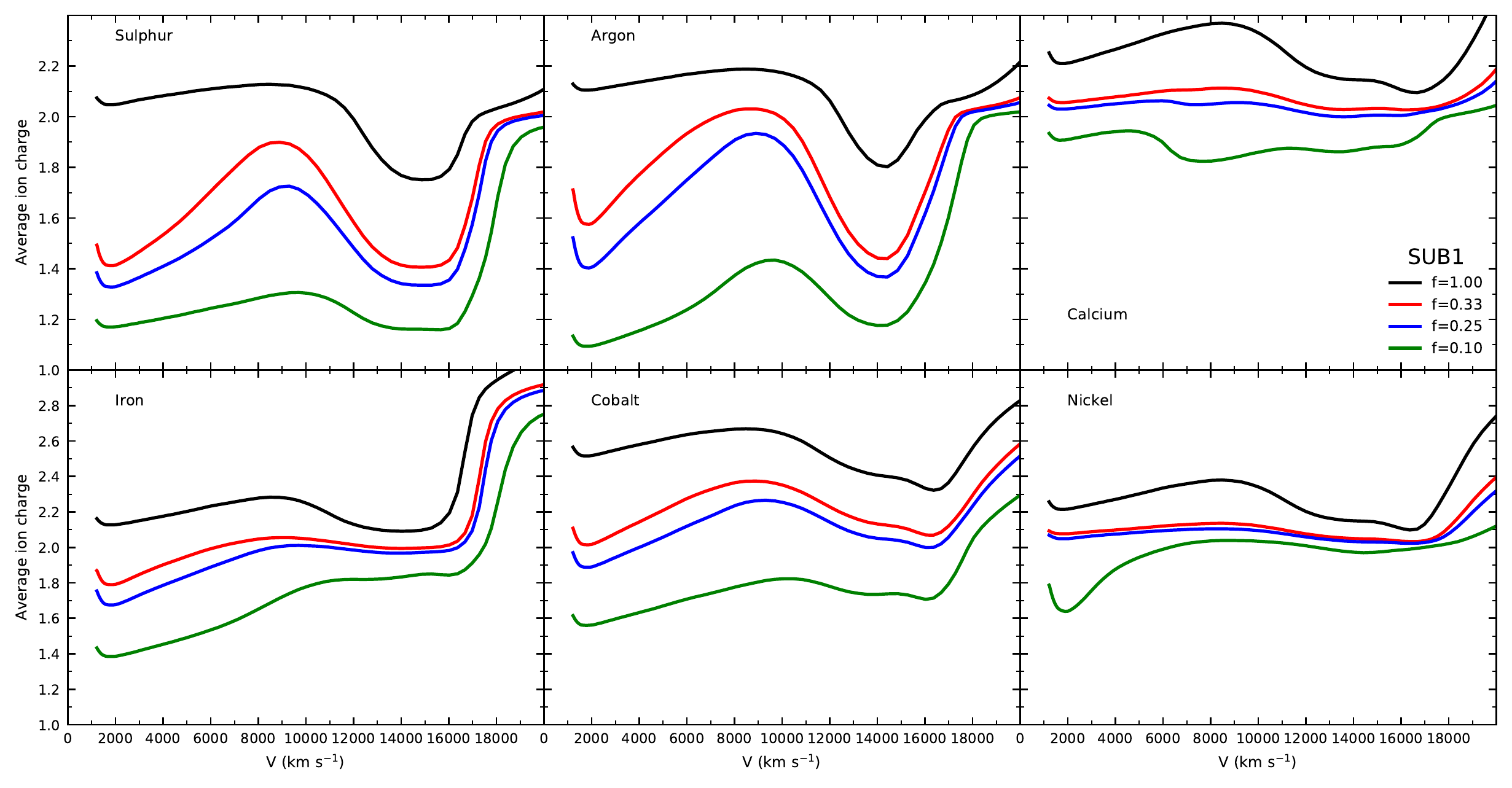}
\end{minipage}
\vspace{-0.5cm}
\caption{Model SUB1 average charge on species for sulfur, argon, calcium, iron, cobalt, and nickel for clumping factors of 1.00, 0.33, 0.25, and 0.10 at roughly $+$200 d after maximum. The ionization is largely sensitive to the clumping factor and the non-thermal energy deposition in the inner ejecta. Sulfur and argon show changes of roughly one electron, while iron shifts from mostly \ions{Fe}{2+} to nearly equal parts \ions{Fe}{2+} and \ions{Fe}{+}. The majority of the stable nickel is located in the inner ejecta, and in this region the ionization is driven by the radioactive decay deposition. These changes in the inner region relate to large changes on the spectra.}
\label{ch4:avg_ion_sub1_figure}
\end{figure}
\end{landscape}

\begin{landscape}
\begin{figure}
\begin{minipage}[t]{\linewidth}
\centering
%\hspace*{-0.5cm}\includegraphics[scale=0.77]{DDC10_plots/avg_ion_all.pdf}
\includegraphics[scale=0.95]{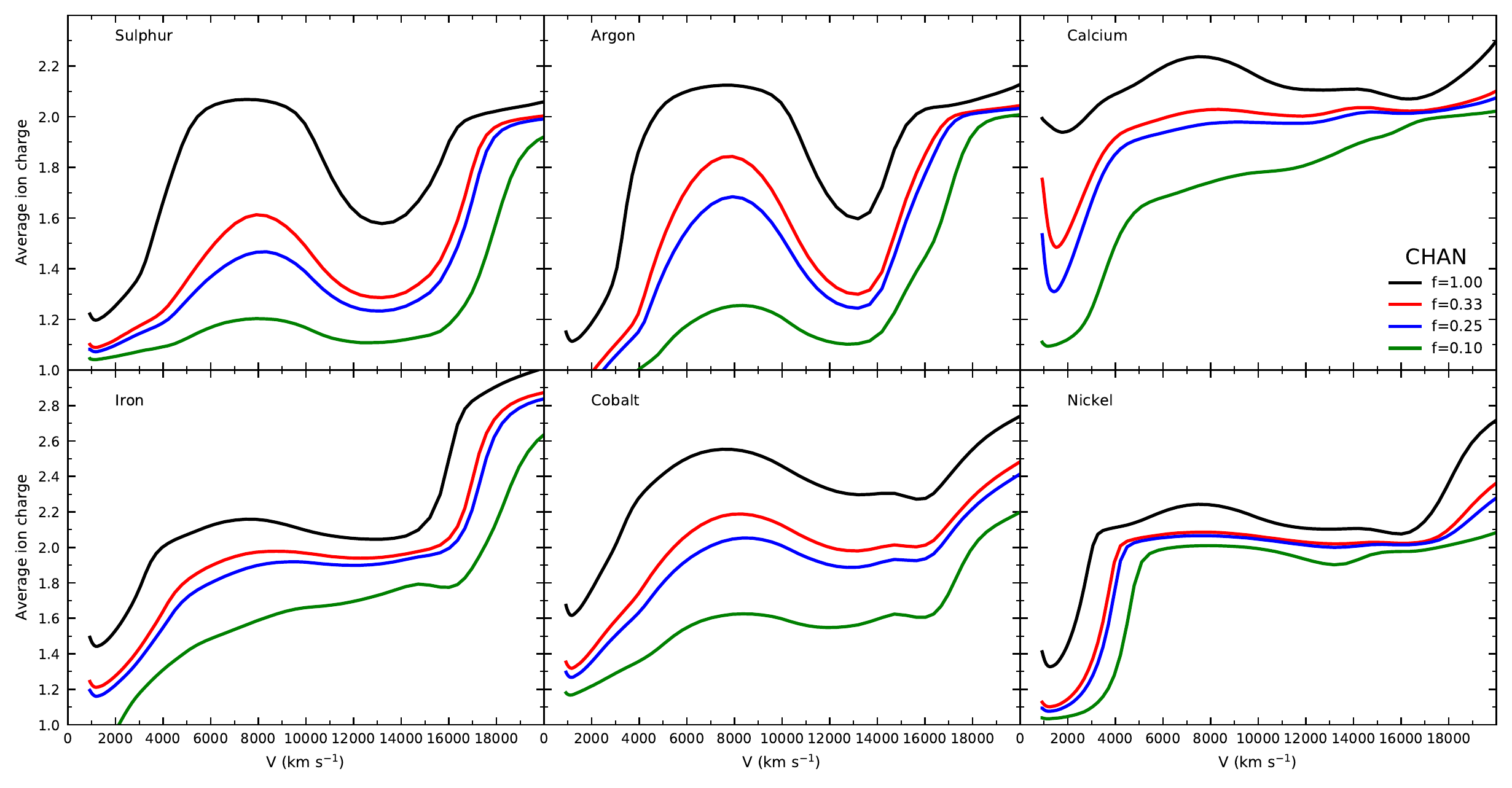}
\end{minipage}
\vspace{-0.5cm}
\caption{Same as \fig~\ref{ch4:avg_ion_sub1_figure} but for model CHAN. The changes in the ionization are smallest (except for calcium) in the region of the ``\iso{Ni}{56} hole'' between different amounts of clumping. The slope of the average ionization of iron and other species in the inner region highlights that small changes can shift the ionization.}
\label{ch4:avg_ion_chan_figure}
\end{figure}
\end{landscape}

%
%88888888888888888888888888888888888888888888888888888888
%
\figs~\ref{ch4:subchan_optical_spec_figure} and \ref{ch4:chan_optical_spec_figure}, show that the \fei, \feii, and \feiii\ optical features changed significantly. The [\feiii] \lb\lb4658, 4702 feature dropped in flux by more than a factor of two in both ejecta models from $f=1.00$ to $f=0.10$, while we saw the emergence of \fei\ features between 4100--4500 \AA\ and between 5400--5600 \AA\ (z $^5$D$^{\rm o}_J$ -- a $^5$F$_{J'}$ optical transitions). \figs~\ref{ch4:subchan_optical_spec_figure}~and~\ref{ch4:chan_optical_spec_figure} show the \fei\ emission as a shoulder to the neighboring [\feiii] and [\feii] emission for $f=$0.33 and 0.25, while we see a noticeable peak for $f=0.10$.

While \feii\ features were generally enhanced, the expected \feii\ feature around 4359 \myunit{\AA} (believed to be the a$^{6}$S$_{5/2}$ -- a$^{6}$D$_{7/2}$ transition and similarly the a$^{6}$S$_{5/2}$ -- a$^{6}$D$_{9/2}$ transition at 4287 \AA) is weak or absent in our models compared with observations. An examination of the individual \fei, \feii, and \feiii\ spectra showed that the optical depth effects are important. 
%It appears that the emergence of \fei\ features caused an optical depth effect with \feii\ and did not produce the strength of the claimed [\feii] \lb\lb4287, 4359 doublet believed to be seen in nebular spectra. 
It appears that the emergence of partially thick \fei\ lines, \tkii\ lines like 4395 \AA, and permitted \feii\ lines reduce the strength of the claimed [\feii] \lb\lb4287, 4359 doublet believed to be seen in nebular spectra. 
%We also see optical depth effects with \tkii\ (4395 \AA) as well as with other permitted \feii\ lines.
%We also find \tkii\ lines (like 4395 \AA) as well as permitted \feii\ lines reduce the strength of this .

\begin{figure}
\centering
\includegraphics[scale=0.80]{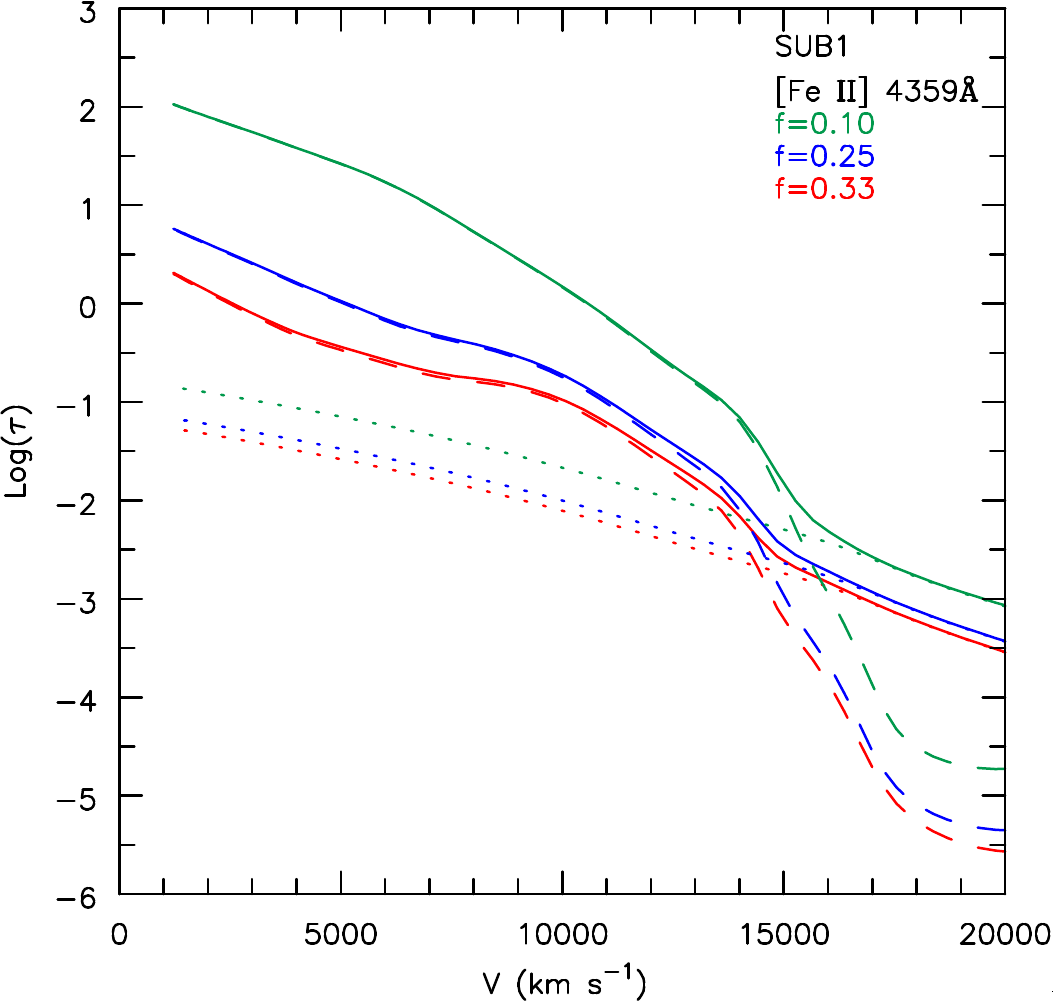}
\caption{The optical depth to [\feii] \lb4359 resonance zone at velocity $V$ for various clumping factors for SUB1. We consider this along the first core impact parameter ($p_{_1}=0$). The dashed lines correspond to the sum total of all lines that interact with the [\feii] \lb4359 line emitted at a particular resonance zone. The dotted lines correspond to the continuum contribution, and the solid lines represent the total [\feii] \lb4359 optical depth to the resonance zone.}
\label{ch4:fig_sub1_fe2_opac}
\end{figure}

\begin{figure}
\centering
\includegraphics[scale=0.80]{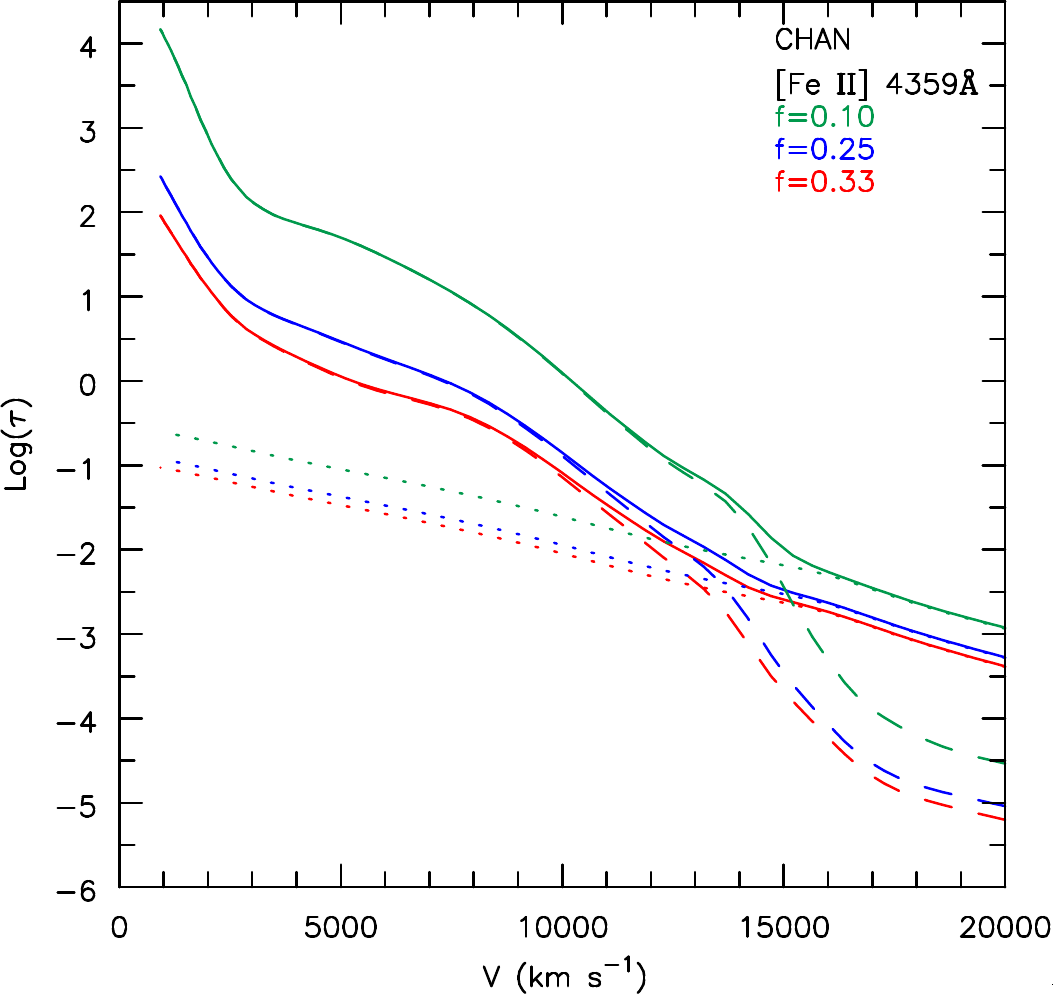}
\caption{Same as \fig~\ref{ch4:fig_sub1_fe2_opac} but now for model CHAN.}
\label{ch4:fig_chan_fe2_opac}
\end{figure}

\begin{landscape}
\begin{figure}
\centering
\includegraphics[scale=0.9]{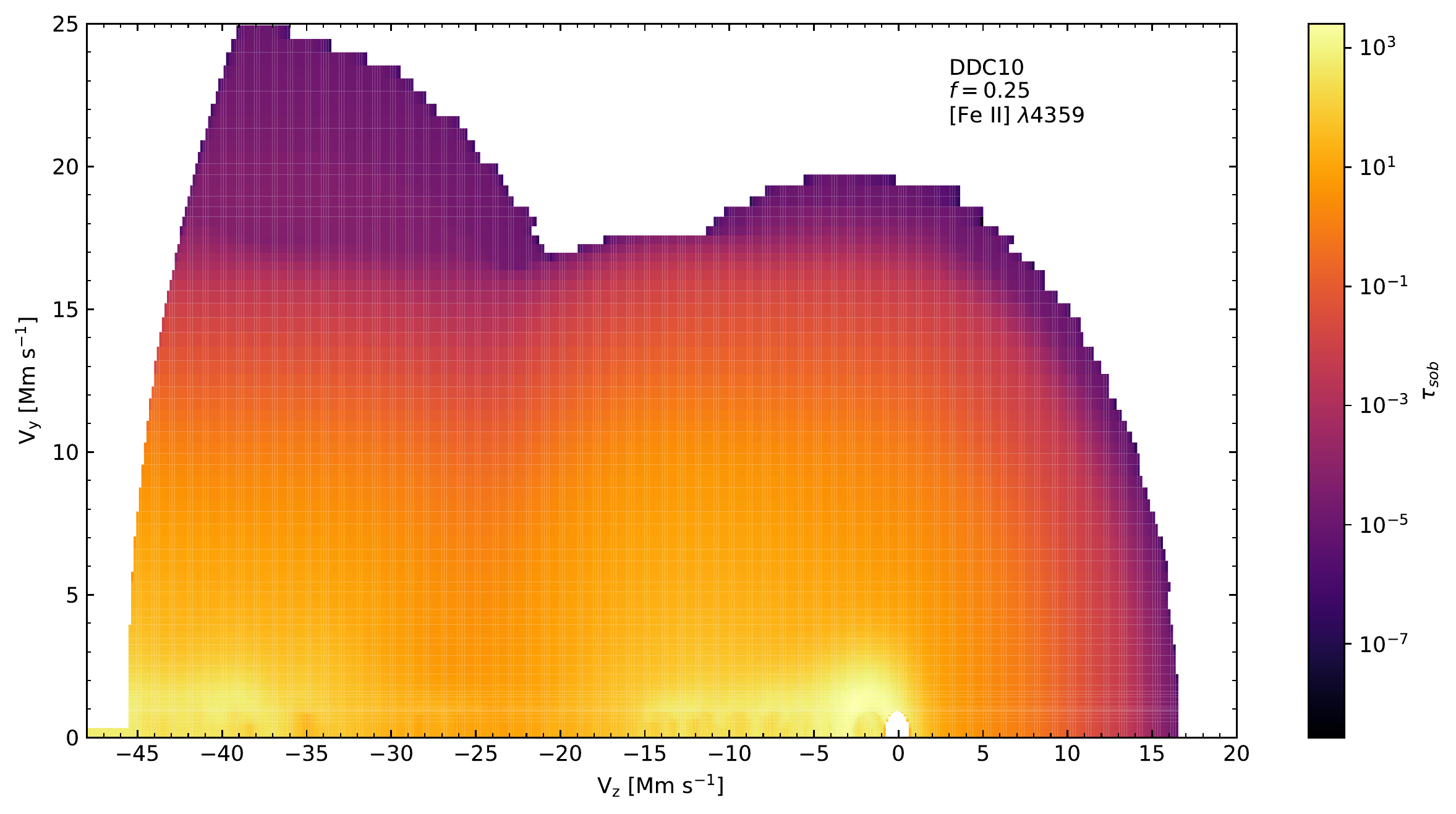}
\caption{The 2D optical depth of the [\feii] \lb4359 (just as in \figs~\ref{ch4:fig_sub1_fe2_opac}~and~\ref{ch4:fig_chan_fe2_opac}) along rays parallel to the $z$-axis for various impact parameters ($y$-direction). We only consider interacting lines whose Sobolev optical depth at their resonance zone is greater than $10^{-5}$. We interpolate the results from our 1D grid onto a rectangular grid. The white space is where the optical depth is zero.}
\label{ch4:taulr_2d_fig}
\end{figure}
\end{landscape}

%In \figs~\ref{ch4:fig_sub1_fe2_opac},~\ref{ch4:fig_chan_fe2_opac}, and~\ref{ch4:taulr_2d_fig} we show the optical depth of the [\feii] \lb4359 line emitted at resonance velocity $V$ and interacting with other lines at other resonance zones. 
In \figs~\ref{ch4:fig_sub1_fe2_opac},~\ref{ch4:fig_chan_fe2_opac}, and~\ref{ch4:taulr_2d_fig} we show the optical depth to the [\feii] \lb4359 resonance zone (at velocity $V$) arising from interactions with other lines along the line of sight. There are tens of interacting lines to the outer ejecta and $\sim$25\,000 interacting lines to the inner ejecta. \figs~\ref{ch4:fig_sub1_fe2_opac}~and~\ref{ch4:fig_chan_fe2_opac} show this optical depth (and the continuum optical depth) along a core ray through the ejecta. The line transitions with the largest optical depth contributions are \fei] \lb4384, \fei] \lb4375, \tkii\ \lb4395, \feii\ \lb4385, and \fei] \lb4404. Differences between models and levels of clumping reflect the differences in the ionization of iron since the largest contributions to the optical depth primarily come from iron lines. \fig~\ref{ch4:taulr_2d_fig} shows this optical depth along various rays parallel to the $z$-axis towards the observer ($V_y$ and $V_z$). In both models, [\feii] \lb4359 reaches an optical depth of unity at roughly 10\,000 \kms\ for a clumping factor of 0.10. In model CHAN, the Sobolev line optical depth reaches unity around 5\,000 \kms\ for $f=0.33$, while for model SUB1 it is true around 2500 \kms. \figs~\ref{ch4:subchan_optical_spec_figure} and~\ref{ch4:chan_optical_spec_figure} highlight how little flux is seen below 4500 \AA. 

\begin{figure*}
\centering
\begin{tabular}{c}
\includegraphics[scale=0.75,angle=-90]{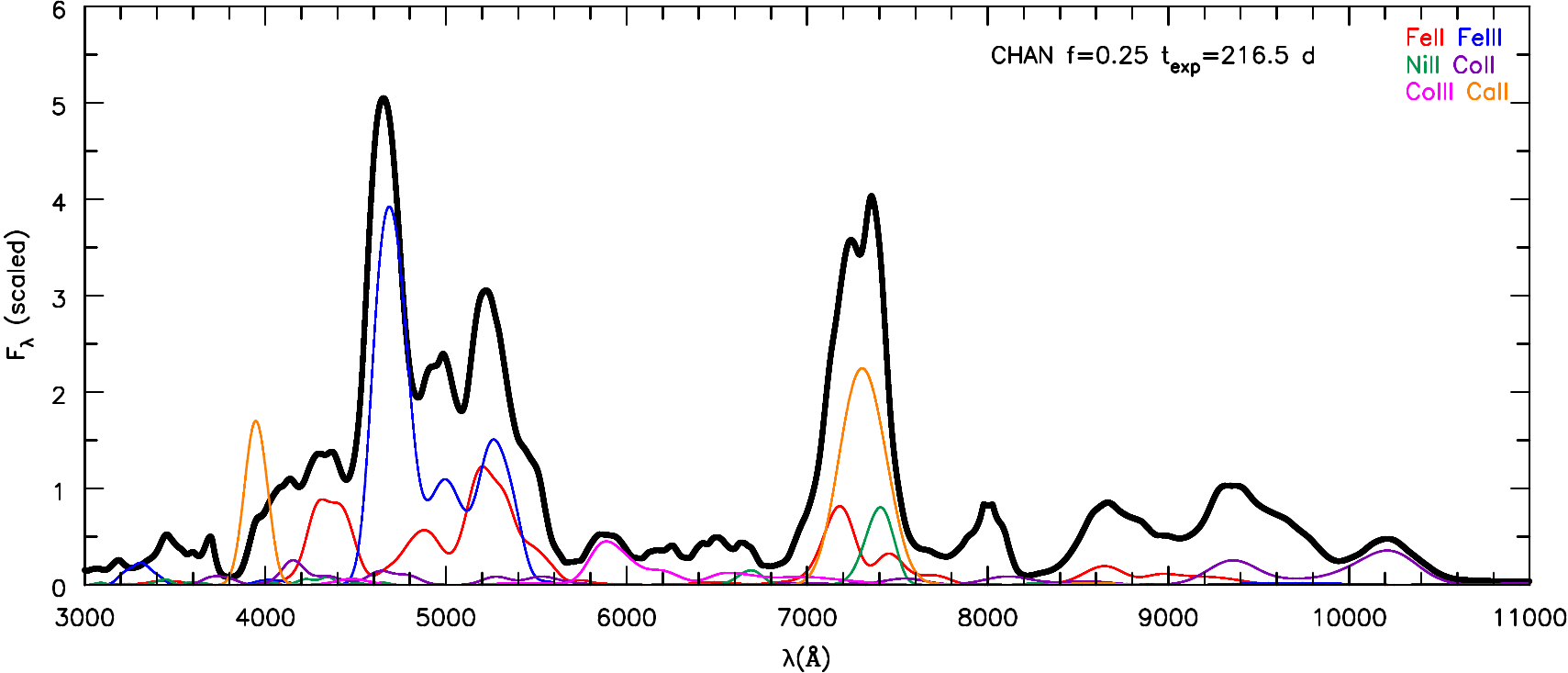}\\
\includegraphics[scale=0.75,angle=-90]{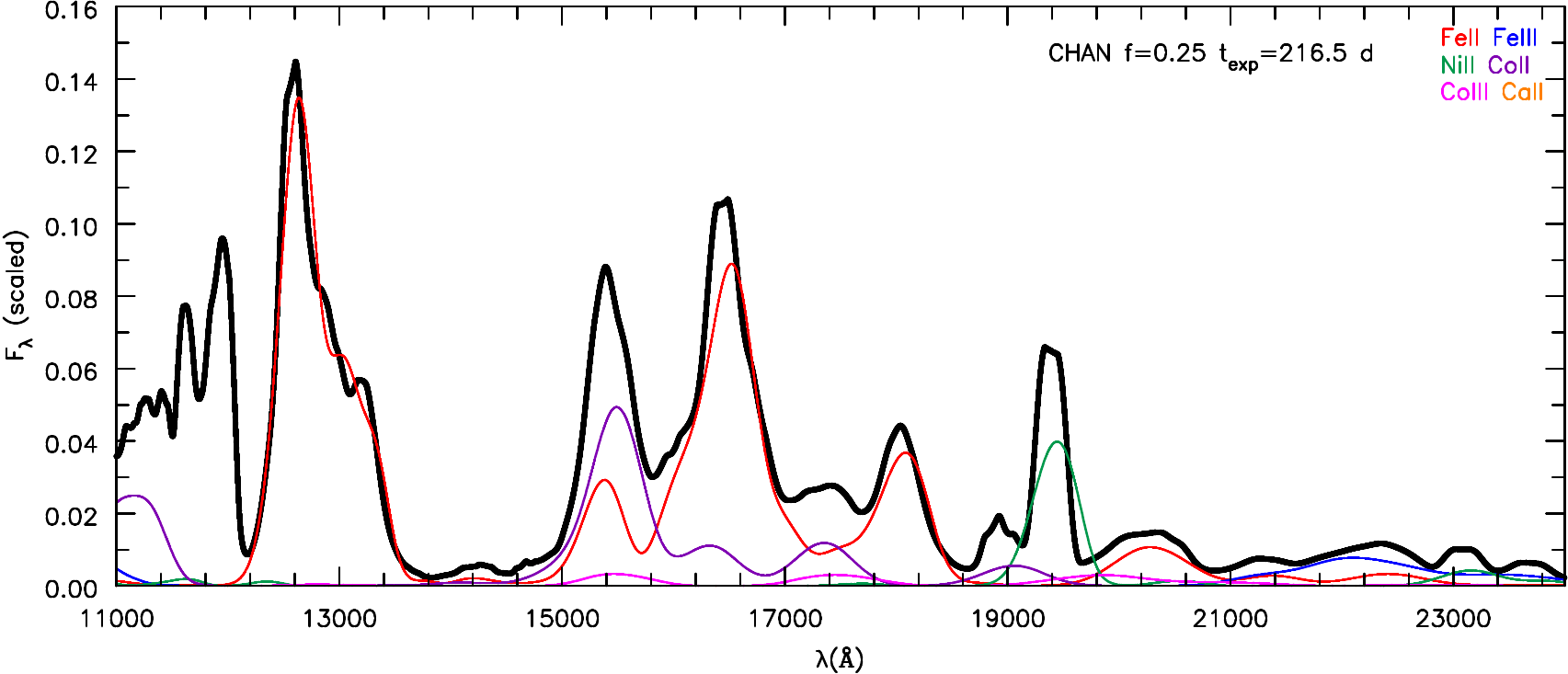}
\end{tabular}
\caption{Observer's frame emission spectra of \feii, \feiii, \nkii, \coii, \coiii, and \caii. We show the full observer's frame spectrum in black. The emission spectra are calculated by taking the temperature and ionization structure output from \cmfgen\ and re-solving the level populations assuming only collisional processes and radiative decays. We assume Gaussian emission profiles of 100 \kms\ in the CMF. }
\label{ch4:emission_spectrum}
\end{figure*}

Studies of SN Ia nebular spectra often investigate the [\feii] feature around 12\,600 \AA\ \citep{Maguire2018,Diamond2018} since it is the least blended feature in nebular SN Ia spectra (free from other lines and ionization states of Fe). Our models confirm it is ``blend''-free. This line complex is therefore the best feature to constrain the \feii\ emitting region. We show in \fig~\ref{ch4:emission_spectrum} that this [\feii] feature can be reproduced by calculating an emission spectrum. To produce this emission spectrum, we use the temperature and ionization structure from \cmfgen\ and re-solve the level populations considering only collisional processes and radiative decays. The \rrte\ is then solved assuming zero opacity. We show this line complex can be inferred from tomography, as it is only sensitive to the temperature and ionization. However, this method will break down for departures from spherical symmetry.

Various observations of nebular SN Ia spectra show a slight shouldering on the 4600 \myunit{\AA} iron complex due to a potential \fei\ feature at 5500 \AA\ -- z $^5$D$^{\rm o}_J$ -- a $^5$F$_{J'}$ optical transitions -- \citep{Childress2015,Graham2017}. Such \fei\ features could constrain the level of ionization within SN ejecta and assist future modelling. Since \fei\ features cause optical depth effects with [\feii] \lb4287, 4359, it is important to determine the Fe ionization. Despite all the observations showing [\feii] \lb4287, 4359, this feature has yet to be accurately modelled by other researchers (Spyromilio 2016, private communication; Sim 2016, private communication). Shifts in the ionization of iron (\ions{Fe}{2+} $\to$ \ions{Fe}{+} $\to$ Fe) are expected as the ejecta continuously expands and cools as less energy is deposited from radioactive decays \citep{Fransson2015}. \cite{Fransson2015} showed that at very late times (1000 days), the 4600 \myunit{\AA} iron complex, despite its similar appearance to early epochs, is dominated by emission from \fei\ and \feii. 
%
%88888888888888888888888888888888888888888888888888888888
%
\subsection{[Ni~{\sc II}] 1.939 microns}\label{ch4:subsection_nickel_line}
%
%88888888888888888888888888888888888888888888888888888888
%
In SUB1 models, the ionization fraction for nickel (Ni$^{+}$/ Ni$^{2+}$) drops in the inner region ($\lesssim 5000$ \kms) by roughly three orders of magnitude when changing $f=1.00$ to $f=0.10$. However, the [\nkii] 1.939 $\mu$m line is still absent in SUB1. Where the line formation occurs, below $\sim$6500 \kms, the electron density is higher than 10$^{7}$ cm$^{-3}$ in the inner part of the ejecta, so its line emission scales linearly with density as we increase the amount of clumping. These densities are above the critical density for the upper level of this transition, so collisional de-excitations are important.

With $f=0.10$ for SUB1, the weak [\nkii] 1.939 micron line is also a factor of a few weaker than [\coii] 1.9035 microns. SUB1 has more than a factor of 2 less stable nickel compared to CHAN (Table~\ref{ch4:model_info_abund}) in which we do still see the [\nkii] 1.939 micron line. With all of these factors such as ionization and abundance accounted for, it is not surprising that spectra of SUB1 still do not show this line.
%
%88888888888888888888888888888888888888888888888888888888
%
\subsection{IMEs}\label{ch4:subsection_IMEs}
%
%88888888888888888888888888888888888888888888888888888888
%
Due to the overlap in production/mixing of \iso{Ni}{56} and IMEs in our models, the strengths of IMEs are sensitive to the non-thermal heating and ionization structure within the ejecta. Within our models, a consequence of clumping is the enhancement of the [\caii] \lb\lb7291, 7324 doublet. The [\caii] doublet is blended with the [\feii] \lb\lb7155, 7172, 7388 triplet and [\ariii] \lb7136, and it contributes a large portion of the flux to the blended feature (see \figs~\ref{ch4:sub1_f0p1_spec_components_figure}-\ref{ch4:chan_f0p33_spec_components_figure}). The [\caii] emission flux is model dependent. Not only is the mass of calcium $\sim$1.75 times larger in CHAN than SUB1 but also the distribution of calcium varies significantly between SUB1 and CHAN. 
%In SUB1, 75 percent of the calcium mass and 20 percent of the total energy deposited is beyond 10\,000 \kms. In CHAN, however, 60 percent of the calcium mass and nearly 85 percent of the total energy deposited is below 10\,000 \kms. These models help constrain the amount of stratification between the original \iso{Ni}{56} and IMEs required to produce SNe Ia nebular spectra.
In SUB1, the inner region containing 80 percent of the energy deposited only contains 25 percent of the calcium mass. In CHAN, however, the inner region containing 80 percent of the energy deposited contains 50 percent of the calcium mass. These models help constrain the amount of stratification between the original \iso{Ni}{56} and IMEs required to produce SNe Ia nebular spectra.

Once the ionization is lowered, Ca$^{+}$ becomes the dominant cooling for the zone rich in IMEs, since S$^{+}$ and Ar$^{+}$, unlike their twice-ionized siblings, do not have strong cooling lines due to their low critical densities. We see the strength in the twice-ionized sulfur and argon lines ([\siii] \lb\lb9068, 9530 and [\ariii] \lb\lb7135, 7751) decreases as we lower the clumping factor. 
%However, it is unclear if we see such strong [\caii] \lb\lb7291,7324 emission blended into this feature around 7200 \AA, which is thought to be mostly [\feii] and [\nkii], in nebular spectra -- \cite{Taubenberger2013,Bikmaev2015,Graham2017,Maguire2018}. For low luminosity 91bg-like SNe Ia (such as SN1999by), modelling suggests Ca emission is dominant component of the 7200 \AA\ feature at +180 days \citep{Blondin2018}. 
However, it is unclear if such strong [\caii] \lb\lb7291, 7324 emission is seen in spectra of classic SNe Ia blended into the feature around 7200 \AA, which is thought to be mostly [\feii] and [\nkii], in nebular spectra -- \cite{Taubenberger2013,Bikmaev2015,Graham2017,Maguire2018}. For low luminosity 91bg-like SNe Ia (such as SN1999by), modelling suggests Ca emission is the dominant component of the 7200 \AA\ feature at +180 days \citep{Blondin2018}.

If the 7200 \AA\ feature is highly blended with [\caii], then for a given electron density and temperature, we can predict emissivity per ion ratio of the [\caii] doublet to the [\feii] lines using some simple physical assumptions. Assuming only collisional processes and radiative decays, we can solve for the atomic level populations for a range of temperatures and electron densities. \fig~\ref{ch4:ca2_on_fe2_emiss_ratio} shows the emissivity ratio of the [\caii] \lb7291 transition to the [\feii] \lb7155 transition for a range of temperatures and electron densities with our simple assumptions.
%What we see is that for equal $N({\rm Fe}^+)/N({\rm Ca}^+)=1$, the emissivity ratio between [\caii] \lb7291 and [\feii] \lb7155 for temperatures between 2000-7000 K and electron densities between 10$^5$-10$^8$ cm$^{-3}$ is between a factor of 10-100 as shown in \fig~\ref{ch4:ca2_on_fe2_emiss_ratio}. 
As shown in \fig~\ref{ch4:ca2_on_fe2_emiss_ratio}, for equal $N({\rm Fe}^+)/N({\rm Ca}^+)=1$, the emissivity ratio between [\caii] \lb7291 and [\feii] \lb7155 for temperatures between 2000-7000 K and electron densities between 10$^5$-10$^8$ cm$^{-3}$ is between a factor of 10-100. 
However, in our models, our $N({\rm Fe}^+)/N({\rm Ca}^+)$ is $\sim$5-30 between 5\,000-10\,000 \kms. Therefore, we expect to see [\caii] \lb\lb7291, 7324 emission blended as long as the $N({\rm Fe}^+)/N({\rm Ca}^+)$ is $\lesssim$ 100. Although this only relates to the stronger component of the [\caii] doublet, the [\caii] \lb7324 line comprises roughly 40 percent of the overall contribution from [\caii] (see Table~\ref{ch4:table_line_transitions}). The other [\feii] blended components will contribute much less flux as the Einstein A values are a factor of 2-3 less than the [\feii] \lb7155 transition. 

Despite the level excitation energy of Ca$^+$ $^2$D$_{5/2}$ and Fe$^+$ a$^2$G$_{9/2}$ being similar ($\sim$16 percent difference), the average level populations in LTE can be several orders of magnitude different for the same total ion population. The average LTE level population compared to the total is simply
\begin{eqnarray}
	\frac{\langle N({\rm Fe}^{+}[{\rm a}\,{}^2{\rm G}_{9/2}])\rangle }{ N({\rm Fe}^{+}) } &=& \frac{ 1 }{ Z }g_{{\rm a}\,{}^2{\rm G}_{9/2}} e^{-E[{\rm a}\,{}^2{\rm G}_{9/2}]/kT} \nonumber \\
	&=& \frac{ 1 }{ Z }10 e^{-1.964\,{\rm eV}/kT},
\end{eqnarray}
where $Z$ is the partition function. For temperatures of 2000, 5000, and 10\,000 K, the partition function (using the first 18 levels) is approximately 28, 43, and 58 respectively assuming the states are in LTE with respect to the ground state. 
%\begin{eqnarray}
%    Z &\approx& 10+8e^{-0.0477\,{\rm eV}\beta}+6e^{- 0.0828\,{\rm eV}\beta}+4e^{-0.107\,{\rm eV}/\beta}+ \nonumber \\ 
%    && 2e^{-0.121\,{\rm eV}\beta}+ 10e^{-0.232\,{\rm eV}\beta}+8e^{-0.301\,{\rm eV}\beta}+\nonumber \\
%    && 6e^{-0.352\,{\rm eV}\beta}+4e^{-0.387\,{\rm eV}\beta} + 8e^{-0.986\,{\rm eV}\beta}+\nonumber \\
%    && 6e^{-1.040\,{\rm eV}\beta}+4e^{-1.076\,{\rm eV}\beta}+ 2e^{-1.097\,{\rm eV}\beta}+ \nonumber \\
%    &&  6e^{-1.671\,{\rm eV}\beta}+4e^{-1.695\,{\rm eV}\beta}+2e^{-1.724\,{\rm eV}\beta} + \nonumber\\
%    && 10e^{-1.964\,{\rm eV}\beta}+8e^{-1.964\,{\rm eV}\beta}+ \ldots
%\end{eqnarray}
%and $\beta=1/kT$.
Since Fe$^+$ has many easily excited lower levels, then even for the same ion abundance between Ca$^+$ and Fe$^+$, the emission from \caii\ will dominate the blended feature at 7200 \AA. 

The strong emission of the [\caii] \lb\lb7291, 7324 doublet is not only a result of the ionization, temperature, and electron density but also coupled to the radiation field. The \caii\ H\&K lines have large oscillator strengths and can pump electrons into the upper $^2$P$^{\rm o}$ levels. They then decay to the $^2$D state, which again decays to the $^{2}$S ground state giving us [\caii] \lb\lb7291, 7324 emission. We have taken our ionization, temperature, and electron density structure from {\cmfgen} and re-solved for the level populations of Ca$^{+}$ assuming only collisional and radiative decay processes. For levels that are coupled to UV transitions, this assumption is only accurate within 50\%. We then solved the \rrte\ for line emission of \caii\ with zero opacity. Our results (see \fig~\ref{ch4:emission_spectrum}) show the spectra of \caii\ is sensitive to the radiation field. Flux is absorbed in the \caii\ H\&K lines which can then be emitted in the \caii\ NIR triplet as well as the [\caii] \lb\lb7291, 7324 doublet. Despite the critical densities for \ions{Ca}{+} $^2$D$_{3/2}$ and $^2$D$_{5/2}$ being on the order of $\sim 10^{5}-10^{6}$ cm$^{-3}$, the atomic levels above the \ions{Ca}{+} $^2$D$_{3/2}$ and $^2$D$_{5/2}$ have sufficiently high critical densities which inhibits collisional de-excitations locking these levels to their LTE value. 

Such strong [\caii] emission could indicate a problem with our atomic data for \fei/\feii\ or likely be an indication of more stratified material in our 1D models. Element stratification within the ejecta also influences the strength of the lines belonging to IMEs. In particular, the presence of some \isoco\ in the IME zone means that positrons are available as a heating source after the ejecta has become optically thin to \gray\ photons. However, even at late times, this contributes only a fraction of non-thermal heating in the inner region. \grays\ still scatter out into the ejecta, heating some of the outer layers.

If the problem lies not with our atomic data, then stratification is the cause of strong IME emission. This puts a constraint on the nucleosynthetic distribution of calcium produced in SNe Ia. This either prohibits an overlap between the calcium and the radioactive isotopes like \isoni, or this constrains the mass of calcium produced during the nucleosynthesis. Not surprisingly, when we artificially scaled down the mass of calcium by a factor of 2 in our CHAN $f=0.25$ clumped model, the peak flux drops by half. This also has implications for understanding the early-time light curves. In order to produce the blue colors of SNe Ia, studies routinely mix \iso{Ni}{56} into the outer layers of the ejecta in DDT models.
%
%One test that we have performed is to artificially scale the mass of calcium down by a factor of 2 in our model CHAN of $f=0.25$ and add the difference to the dominant species at each depth. When we did this test, what we saw was only a reduction of the peak flux by {\red nearly 50 percent}.
%
\section{Conclusion}\label{ch4:section_conclusion}%%%%%%%%%%%%%%%%%%%%
We have performed 1D radiative transfer calculations using \cmfgen\ for two ejecta models (one sub-\Mch\ detonation and one \Mch\ DDT -- see \cite{Wilk2018} for more details) utilizing micro-clumping at 216.5 days post explosion.
%We aimed to understand nebular spectra by comparing the effects of micro-clumping against un-clumped spectra, which seem to show a higher ionization spectrum that what is observed. 
Our goal was to understand the influence of micro-clumping on nebular spectra and to test when clumped models would provide a better fit to the observed level of ionization in Ia spectra.
Clumping is expected to occur naturally in SN ejecta and naturally reduces the ionization by enhancing recombination. We considered three different clumping factors, 0.33, 0.25, and 0.10, which we assumed were constant throughout the ejecta. Our models are in a regime where potentially small changes can shift the ionization. This also means that our models are sensitive to atomic physics. 

%Clumping lowers the average ionization of all species, notably for IMEs and IGEs. The average ionization of IMEs is reduced by about one electron below 10\,000 \kms and roughly one half of an electron for IGEs (except cobalt which is reduced by roughly an electron). Despite model SUB1 lowering its ionization with clumping, the nickel ionization remained high in the inner region due to a lack of a ``\iso{Ni}{56} hole.'' With an already reduced stable nickel mass compared to CHAN, SUB1 failed to show strong emission from \nkii\ in the optical and NIR with the [\nkii] 1.939 \mum\ line.
Clumping lowers the average ionization of all species. The average ionization of IMEs is reduced by about one electron below 10\,000 \kms\ and roughly one half of an electron for IGEs (except cobalt which is reduced by roughly an electron). Despite clumping lowering the ionization in SUB1, the nickel ionization remained high in the inner region due to a lack of a ``\iso{Ni}{56} hole.'' With an already reduced stable nickel mass compared to CHAN, SUB1 failed to show strong emission from \nkii\ in the optical and in the NIR with the [\nkii] 1.939 \mum\ line.

As iron is the most abundant species in the inner ejecta at 216 days, clumping had the most visible effect on the flux of [\feiii] features such as [\feiii] \lb\lb4658, 4702. As the iron ionization is lowered, the flux in \feii\ features increases, and permitted \fei\ lines near 5500 \AA\ emerged. These \fei\ lines cause a shoulder to form on the [\feii] and [\feiii] optical blend between 4200-5400 \AA. Our attempts to model the [\feii] \lb\lb4287, 4359 feature were unsuccessful. While clumping generally enhanced the strength of [\feii] features, absorption by other transitions limit its strength. 
%and we show its absence is a result of optical depth effects from other lines. 
Despite the difficulty in reproducing [\feii] \lb\lb4287, 4359, we show that the [\feii] complex around 12500 \AA\ is completely reproduced under simple physics assumptions of collisional and radiative decays from a given ionization and temperature structure. In the IR, particularly from 10\,000-11\,800 \AA, changes in the flux of \fei\ lines of several orders of magnitude for a factor of a few change in density. 

For the same \iso{Ni}{56} mass, \Mch\ explosions produce more IMEs compared to sub-\Mch\ explosions. We have seen that only large clumping can sufficiently suppress emission from IMEs such as \ariii\ and \siii. However, as we increase clumping, both models show an increase in the [\caii] \lb\lb 7291, 7324, which dominates over the [\feii] \lb7155 feature. Since the presence/absence of [\caii] in this 7200 \AA\ blend is still highly uncertain, we suggest that SN Ia ejecta require less mixing between the original \iso{Ni}{56} and the calcium distribution. 
Another possibility is that the Ni and Ca are not microscopically mixed. In that case non-thermal energy deposited in the Fe region would be radiated by Fe, while the energy deposited in the IME region would be radiated primarily by IMEs. In this case the strength of the [\caii] lines relative to [\feii] would be set by the amount of heating in each region. 

Despite arguments that mixing is required to reproduce early-time LCs, mixing between layers of IMEs and \iso{Ni}{56} is inconsistent with what is observed at nebular times, since the emission reflects where most of the energy is deposited. 
%Therefore, evidence of IMEs in nebular spectra is a crucial diagnostic of SNe Ia.
The strength of time-dependent IME features in nebular spectra is a crucial diagnostic for understanding both progenitors and explosion properties of SNe Ia.

Better atomic data can also assist determining the sensitivity it has on the \fei/\feii/\feiii\ features in producing nebular spectra. Because the focus of this work is on nebular phase modelling, further work is necessary to test the time-dependent effects of various levels of clumping. It would be helpful to perform full time-series calculations. Clumping should be fully explored to truly understand the nature of the progenitors of SNe Ia.

\begin{table*}
\centering
\footnotesize
 \begin{tabular}{ l c c c c c c c c } 
 \hline
 Model & Mass & S & Ar & Ca & Fe & Co & \iso{Ni}{58} + \iso{Ni}{60} & \iso{Ni}{56}  \\ [0.5ex] 
 & (\Msun) & (\Msun) & (\Msun) & (\Msun) & (\Msun) & (\Msun) & (\Msun) & (\Msun) \\[0.5ex] 
 \hline
 SUB1 & 1.04 & 1.046(-1) & 2.273(-2) & 2.361(-2) & 2.226(-2) & 5.526(-2) & 1.113(-2) & 5.684(-1) \\ [0.5ex]
 CHAN & 1.40 & 1.661(-1) & 3.693(-2) & 4.120(-2) & 1.020(-1) & 5.713(-2) & 2.517(-2) & 5.708(-1) \\
 \hline
 \end{tabular}
 \normalsize
\caption{Model mass information  in \Msun\ at 0.75 days post explosion. The parentheses (\#) correspond to $\times10^{\#}$.}
\label{ch4:model_info_abund}
\end{table*}

\begin{table*}
    \begin{minipage}[t]{\linewidth}
    \centering
    \small
    \begin{tabular}{c c c c c c c c c c c}
    \hline
        \multicolumn{11}{c}{Ca$^+$} \\ \hline
%        Upper term  (stat. weight) & Lower term (stat. weight) & $A_{ul}$ (s$^{-1}$) & Wavelength (\AA) \\ \hline 
        \multicolumn{3}{c}{Upper level $u$} & \multicolumn{3}{c}{Lower level $l$} & $A_{ul}$ & Wavelength & $N_{\rm crit}(T_4=0.1)$ & $N_{\rm crit}(T_4=0.5)$ & $N_{\rm crit}(T_4=1.0)$ \\ %\hline 
        Term & $g$ & $E_u$ (eV) & Term & $g$ & $E_l$ (eV) & (s$^{-1}$) &  (\AA) & (cm$^{-3}$) & (cm$^{-3}$) & (cm$^{-3}$) \\ \hline 
        $^2$D$_{5/2}$ & 6 & 1.699932 & $^2$S$_{1/2}$ & 2 & 0.000 & 8.025(-1) & 7291.469 & 7.722(5) & 1.717(6) & 2.125(6) \\ \hline
        $^2$D$_{3/2}$ & 4 & 1.692408 & $^2$S$_{1/2}$  & 2 & 0.000 & 7.954(-1) & 7323.888 & 6.070(5) & 1.290(6) & 1.645(6) \\ \hline
        \multicolumn{11}{c}{Fe$^+$} \\ \hline
        \multicolumn{3}{c}{Upper level $u$} & \multicolumn{3}{c}{Lower level $l$} & $A_{ul}$ & Wavelength & $N_{\rm crit}(T_4=0.1)$ & $N_{\rm crit}(T_4=0.5)$ & $N_{\rm crit}(T_4=1.0)$ \\ %\hline 
        Term & $g$ & $E_u$ (eV) & Term & $g$ & $E_l$ (eV) & (s$^{-1}$) &  (\AA) & (cm$^{-3}$) & (cm$^{-3}$) & (cm$^{-3}$) \\ \hline
        a $^2$G$_{9/2}$ & 10 & 1.96448603 & a $^4$F$_{9/2}$ & 10 & 0.23217278 & 1.4950(-1) & 7155.157 & 5.165(6) & 9.130(6) & 9.618(6)\\ \hline
        a $^2$G$_{7/2}$ & 8 & 2.02954814 & a $^4$F$_{7/2}$ & 10 & 0.30129857 & 5.6950(-2) & 7172.004 & 2.415(6) & 4.010(6) & 4.207(6) \\ \hline
        a $^2$G$_{7/2}$ & 8 & 2.02954814 & a $^4$F$_{5/2}$ & 6 & 0.35186476 & 4.3450(-2) & 7388.178 & 2.415(6) & 4.010(6) & 4.207(6) \\ \hline
        a $^2$G$_{9/2}$ & 10 & 1.96448603 & a $^4$F$_{7/2}$ & 10 & 0.30129857 & 4.8450(-2) & 7452.538 & 5.165(6) & 9.130(6) & 9.618(6) \\ \hline
    \end{tabular}
    \normalsize
    \caption{The parentheses (\#) correspond to $\times10^{\#}$. $T_4$ is just $T/(10^4 K)$}
    \label{ch4:table_line_transitions}
    \end{minipage}
\end{table*}


\begin{figure*}
    \centering
    \begin{tabular}{c c}
    \parbox[c]{8.75cm}{\includegraphics[scale=0.65,angle=-90]{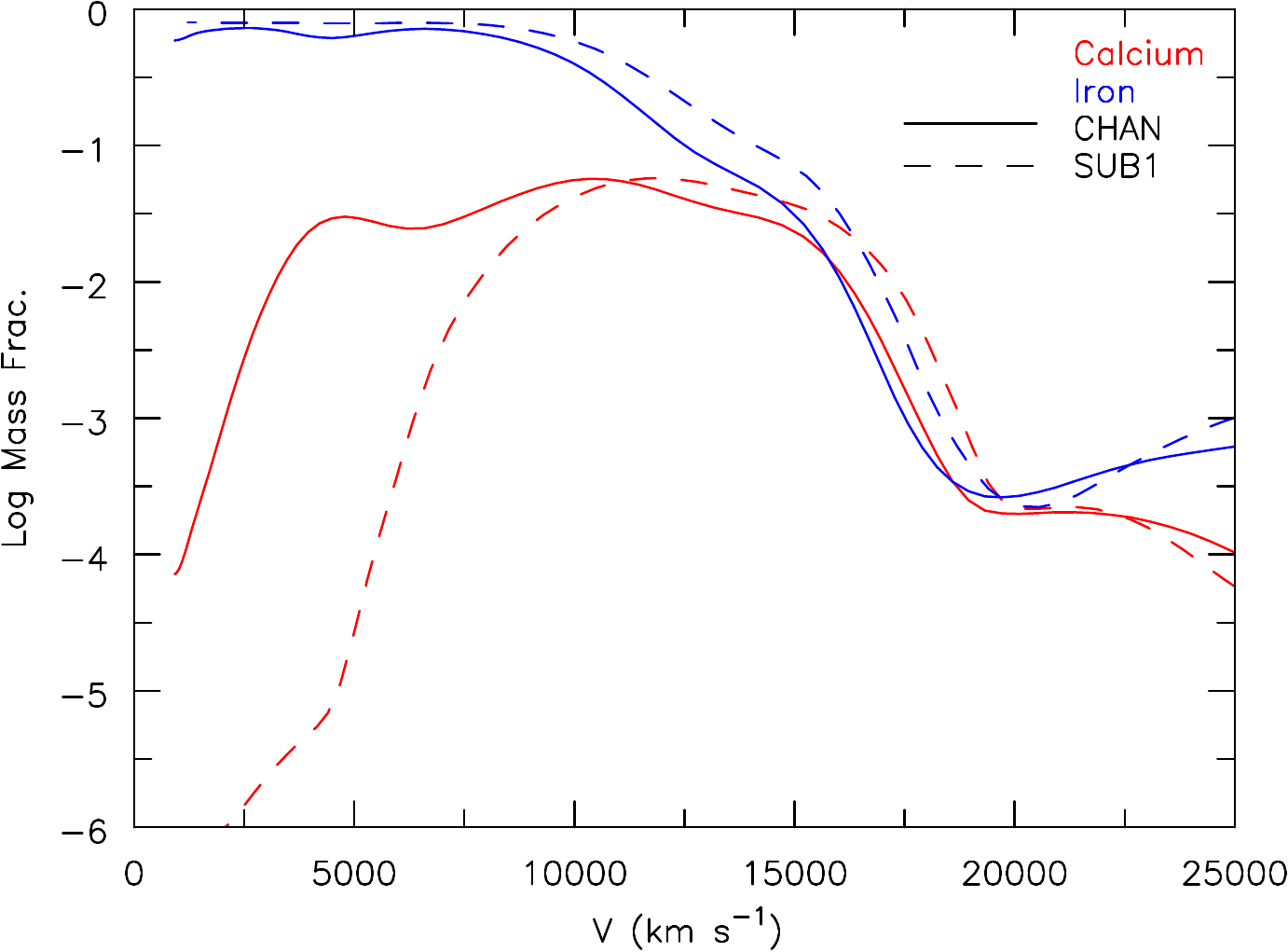}} &     \parbox[c]{8.75cm}{\includegraphics[scale=0.75]{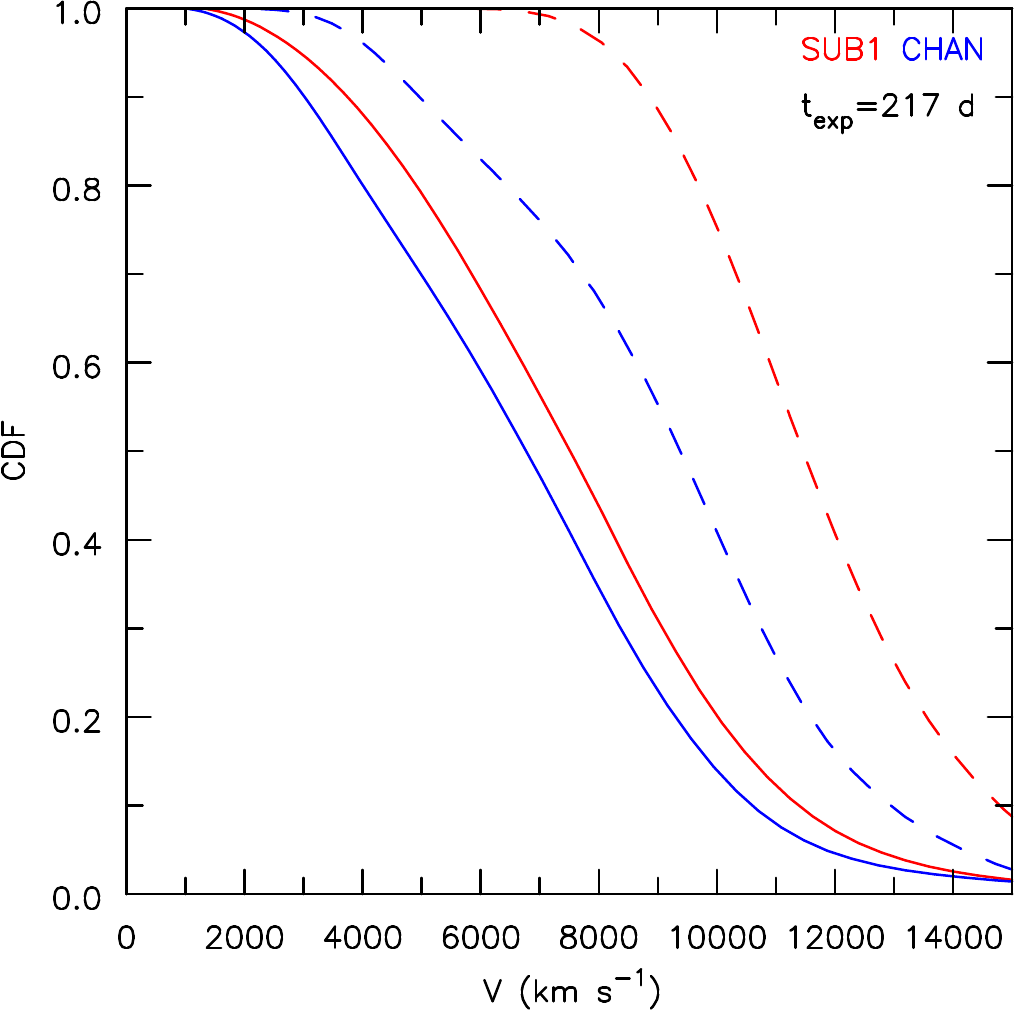}}
    \end{tabular}
    \caption{Left plot is the mass fractions for calcium and iron in both models. The solid line corresponds to model CHAN, and the dashed line corresponds to SUB1. The right plot is the cumulative distribution function for the energy deposition (solid lines) and the calcium mass (dashed lines) at 217 days after explosion (roughly +200 days).}
    \label{ch4:figure_cdf}
\end{figure*}

\begin{figure}
\centering
%\hspace*{-1cm}\includegraphics[scale=0.70]{Figures/other_plots/Ca2_1_3_Fe2_6_17_T_VS_ED.pdf}
\hspace*{-0.5cm}\includegraphics[scale=0.65]{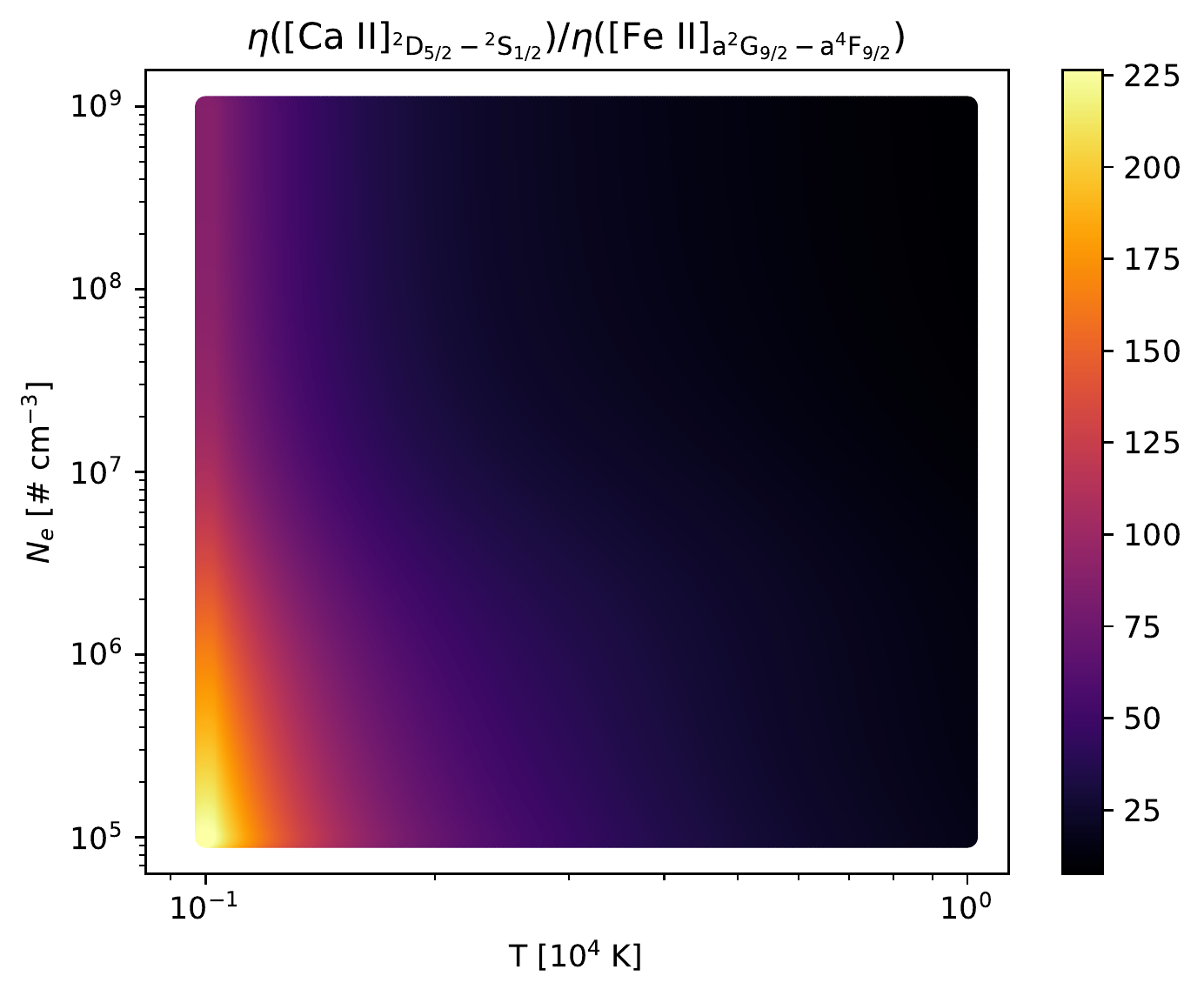}
\vspace*{-0.5cm}
\caption{Line emission ratio of the [\caii] \lb7291 to [\feii] \lb7155 assuming N(\ions{Fe}{+})/N(\ions{Ca}{+})=1. The emission ratio is calculated by solving the rate equations with only collisional and radiative decay terms for a fixed ionization of N(\ions{Fe}{+})/N(\ions{Ca}{+})=1. The emission ratio favors strong calcium. When this ionization ratio is $\gtrsim$50 the iron would begin to dominate the 7200 \AA\ complex.}
\label{ch4:ca2_on_fe2_emiss_ratio}
\end{figure}

%%%%%%%%%%%%%%%%%%%%%%%%%%%%%%%%%%%%%%%%%%%%%%%%%%

%%%%%%%%%%%%%%%%%%%% REFERENCES %%%%%%%%%%%%%%%%%%

% The best way to enter references is to use BibTeX:

\bibliographystyle{mnras}
%\bibliography{bibliography} % if your bibtex file is called example.bib
\input{bibliography.bbl}

% Alternatively you could enter them by hand, like this:
% This method is tedious and prone to error if you have lots of references

%%%%%%%%%%%%%%%%%%%%%%%%%%%%%%%%%%%%%%%%%%%%%%%%%%

% Don't change these lines
\bsp	% typesetting comment
\label{lastpage}
\end{document}